\documentclass{elsarticle}

\usepackage{lineno,hyperref}
\usepackage{color,xcolor}
\usepackage{amsmath}
\usepackage{hhline}

\usepackage{mathrsfs}
\usepackage{amssymb}            
\usepackage[english]{babel}
\usepackage[T1]{fontenc}
\usepackage[latin1]{inputenc}

\usepackage{rotating}

\begin{document}

\begin{frontmatter}

\title{Science orbits in the Saturn-Enceladus circular restricted three-body problem with oblate primaries}

\author[UAE]{Francisco Salazar}
\ead{e7940@hotmail.com}
\author[UAE]{Adham Alkhaja}
\ead{100049385@ku.ac.ae}
\cortext[mycorrespondingauthor]{Corresponding author}
\author[UAE]{Elena Fantino\corref{mycorrespondingauthor}}
\ead{elena.fantino@ku.ac.ae}
\author[IFAC,IMATI]{Elisa Maria Alessi}
\ead{elisamaria.alessi@cnr.it}

\address[UAE]{Department of Aerospace Engineering, Khalifa University of Science and Technology, P.O. Box 127788, Abu Dhabi, United Arab Emirates}
\address[IFAC]{Istituto di Fisica Applicata "Nello Carrara", Consiglio Nazionale delle Ricerche, Via Madonna del Piano 10, 50019 Sesto Fiorentino (FI), Italy}
\address[IMATI]{Istituto di Matematica Applicata e Tecnologie Informatiche  "Enrico Magenes", Consiglio Nazionale delle Ricerche, Via Alfonso Corti 12, 20133 Milano, Italy}

\begin{abstract}
This contribution investigates the properties of a category of orbits around Enceladus. The motivation is the interest in the \textit{in situ} exploration of this moon following Cassini's detection of plumes of water and organic compounds close to its south pole.
In a previous investigation, a set of heteroclinic connections were designed between halo orbits around the equilibrium points $L_1$ and $L_2$ of the circular restricted three-body problem with Saturn and Enceladus as primaries. The kinematical and geometrical characteristics of those trajectories makes them good candidates
as science orbits for the extended observation of the surface of Enceladus: they are highly inclined, they approach the moon and they are maneuver-free. However, the low heights above the surface and the strong perturbing effect of Saturn impose a more careful look at their dynamics, in particular regarding the influence of the polar flattening of the primaries. Therefore, those solutions are here reconsidered by employing a dynamical model that includes the effect of the oblateness of Saturn and Enceladus, individually and in combination. The dynamical equivalents of the halo orbits around the equilibrium points $L_1$ and $L_2$ and their stable and unstable hyperbolic invariant manifolds are obtained in the perturbed models, and maneuver-free heteroclinic connections are identified in the new framework. A systematic comparison with the corresponding solutions of the unperturbed problem shows that qualitative and quantitative features are not significantly altered when the oblateness of the primaries is taken into account. Furthermore, it is found that the $J_2$ coefficient of Saturn plays a larger role than that of Enceladus.
From a mission perspective, the results confirm the scientific value of the solutions obtained in the classical circular restricted three-body problem and suggests that this simpler model can be used in a preliminary feasibility analysis.
\end{abstract}


\end{frontmatter}

\section{Introduction}
The discovery of geyser-like jets of water vapors and organic compounds in the southern polar region of Saturn's moon Enceladus has assigned high priority to this body among exploratory plans in the framework of investigations on life and habitability in other worlds \cite{Porco:2006, Spencer:2006, Parkinson:2007, McKay:2008, MacKenzie:2016, Khawaja:2019}. Future projects could include probes flying through the vapor plumes to collect and analyse their complex organic constituents. 
In order to reach or adequately observe the south polar region of Enceladus, science orbits must have high inclination, low altitude and preferably low eccentricity. However, the design of such orbits around a planetary satellite like Enceladus is challenging due to the intense gravitational perturbation of Saturn. 
Past studies have identified long-term stable orbits around Europa and Enceladus through averaging techniques in the Jupiter-Europa and Saturn-Enceladus Hill's problems \cite{Paskowitz:2006a, Lara:2010}. Global grid searches performed in the Jupiter-Europa circular restricted three-body problem (CR3BP) and in the Saturn-Enceladus unaveraged perturbed Hill's model have disclosed stable periodic solutions in a linear sense about Europa and Enceladus \cite{Lara:2010, Russell:2006}. On the other hand, instead of using complex global grid searches or resorting to doubly-averaged techniques, other investigations have focused on strategies employing Poincar\'{e} sections in the CR3BP to compute trajectories around a planetary moon \cite{Koon:2000, Koon:2011, Gomez:2000, Gomez:2004, Barrabes:2009, Canalias:2006, Haapala:2014}. This means searching for intersections at an appropriate surface of section between stable and unstable hyperbolic invariant manifolds (HIMs) associated with periodic Libration Point Orbits (LPOs). For example, \cite{Davis:2018, Salazar:2019a, Salazar:2019b, Fantino:2020, Salazar:2021} have computed heteroclinic and homoclinic transfers between LPOs, with the velocity difference at the patch point  between the connecting trajectories
representing the cost of the solution. In particular, \cite{Davis:2018} have recently designed a heteroclinic transfer connecting $L_1$ and $L_2$ halo orbits in the Saturn-Enceladus CR3BP. Because of the large out-of-plane component of these LPOs \cite{Breakwell:1979, Howell:1984, Howell:1997}, the solution reaches very low latitudes below the lunar equator. \cite{Salazar:2019b, Fantino:2020} have confirmed and extended this result by constructing connections between Northern halo orbits and Southern halo orbits at $L_1$ and $L_2$. Furthermore, the analysis of the performance of these solutions as science orbits shows that long, uninterrupted, low-altitude views of the polar regions are possible with negligible amounts of fuel. 

Although the CR3BP accounts for the gravitational attraction of all relevant bodies in the study of the motion of a spacecraft performing a mission at Enceladus, approximating Saturn and Enceladus with point masses may be inaccurate when the probe reaches and maintains very low altitudes above the surface of the moon and at the same time stays also quite close to the planet. As a matter of fact, \cite{Bury:2018} has studied the effect of the oblateness of one or both primaries in the Saturn-Enceladus CR3BP and has observed that it disrupts the periodic orbits obtained in the unperturbed CR3BP, causing rapid deviations that may even lead to impacting Enceladus or escaping the system. In recent years, significant progress has been made in the study of the CR3BP with oblate primaries. For example, \cite{Mittal:2009, Mittal:2020} have analysed the variations in shape and energy in a certain family of periodic orbits in the planar CR3BP when one or both primaries are oblate bodies. \cite{Arredondo:2012, Abouelmagd:2012,  Abouelmagd:2015} have studied the effects of the oblateness parameters on the position and linear stability of the libration points in the planar circular restricted three-body problem. \cite{Safiya:2012} has explored the effect of the oblateness of Saturn on planar periodic and quasi-periodic orbits around both primaries in the Saturn-Titan CR3BP. \cite{Bury:2018} has studied the Jupiter-Europa and Saturn-Enceladus systems and has derived the equations of motion for the spatial CR3BP in the presence of the $J_2$, $J_3$, and $J_4$ zonal harmonic terms of the planet and has analysed their effect on the positions of the equilibria and the dynamical behavior of planar periodic orbits about $L_1$ and $L_2$. 

The present contribution builds upon all the above theories and findings and addresses in a systematic way the effects of the second zonal harmonic term ($J_2$) of both Saturn and Enceladus on the positions of the equilibrium points,  the shape and location of the halo orbits around $L_1$ and $L_2$ and the heteroclinic connections between such orbits, as computed by \cite{Salazar:2019b, Fantino:2020} in the unperturbed CR3BP. Furthermore, the mean motion of the primaries (here assumed in circular orbits) is corrected to take into account the oblateness effects.
The work aims, on the one hand, at refining the results of \cite{Salazar:2019b, Fantino:2020} and, on the other, at discussing their validity as approximate solutions. The analysis is carried out by introducing the perturbation of the oblateness of the two bodies separately and in combination, thus obtaining a quantitative insight into their individual effects.   

Since the heteroclinic connections obtained develop in 3D, they can be proposed as science orbits for an \textit{in situ} mission at Enceladus. 
The periodic character of the halo orbits can be exploited to construct a fuel-efficient exploration tour of this moon made of chains of itineraries in which the departure and arrival halo orbits are used to park the spacecraft between consecutive flights. The inclusion of the oblateness effects of Saturn and Enceladus adds robustness to the solutions and makes them suitable to an operational scenario. 

From a mission perspective, the study of the observational performance of these trajectories is crucial. Kinematical and geometrical parameters such as transfer times, distances from the surface, speeds relative to an Enceladus-centered inertial frame, times of overflight and surface coverage parameters are computed, analysed  
and compared with the equivalent features obtained in the classical (unperturbed) CR3BP.
A preliminary version of this work has been presented at the 71$^{st}$ edition of the International Astronautical Congress 
\cite{Alkhaja:2020}.

\section{The dynamical model}
\label{sec:model}
\begin{figure}[h!]
\centering
\includegraphics[scale = 0.3]{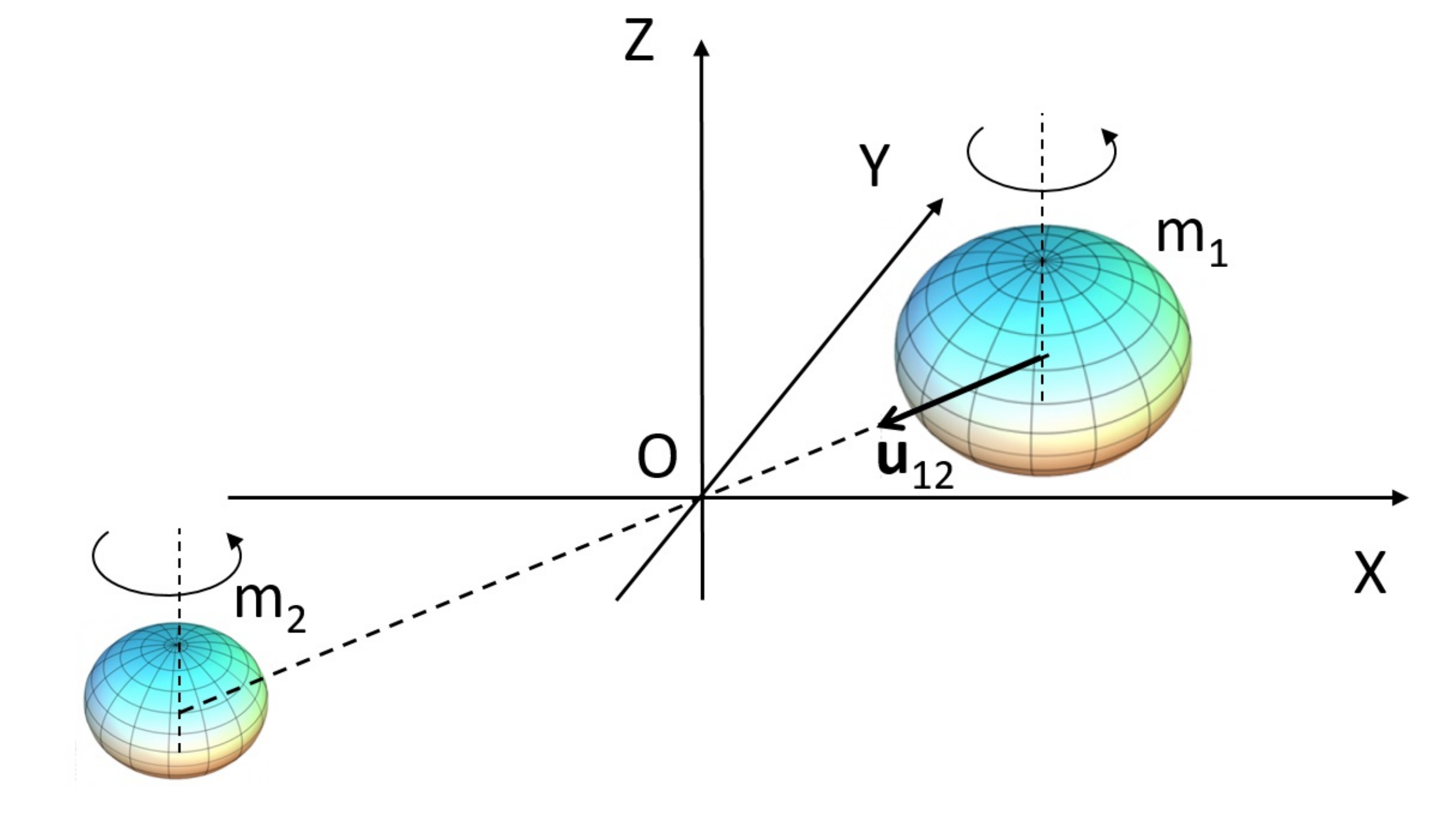}
\caption{Sketch of the system composed by two oblate homogeneous spheroids with the maximum inertia axes parallel to the $Z$-axis of an inertial reference frame centered at their center of mass.}
\label{fig:oblate}
\end{figure}

Let $X$, $Y$ and $Z$ be the position coordinates of a body of negligible mass with respect to an inertial reference frame with origin at the center of mass $O$ of two homogeneous oblate spheroids called primaries, whose equators are contained in the $XY$-plane (Fig.~\ref{fig:oblate}). The external gravitational potential of the system can be written as (see also \cite{Danby:1992}, Chapter 6)
\begin{eqnarray}
\label{eq:V}
V & = & \displaystyle\frac{Gm_1}{r_1}\left\{1-\frac{J_2^1}{2}\left(\frac{R_1}{r_1}\right)^2\left[3\left(\frac{Z}{r_1}\right)^2-1\right]\right\} \nonumber \\
  & + & \displaystyle\frac{Gm_2}{r_2}\left\{1-\frac{J_2^2}{2}\left(\frac{R_2}{r_2}\right)^2\left[3\left(\frac{Z}{r_2}\right)^2-1\right]\right\}, 
\end{eqnarray}
where $G$ is the universal gravitational constant, $m_1$ and $m_2$ are the masses of the primaries, $R_1$ and $R_2$ are their equatorial radii, $r_1$ and $r_2$ denote the distances of the third body from $m_1$ and $m_2$, respectively, and $J_2^i$ ($i$ = 1, 2) is a function of the difference in the principal moments of inertia relative to the north-south axis and an equatorial axis (superscripts 1 and 2 refer to the larger and smaller primary, respectively). For oblate bodies, this coefficient is positive.
\begin{figure}[h!]
\centering
\includegraphics[scale = 0.275]{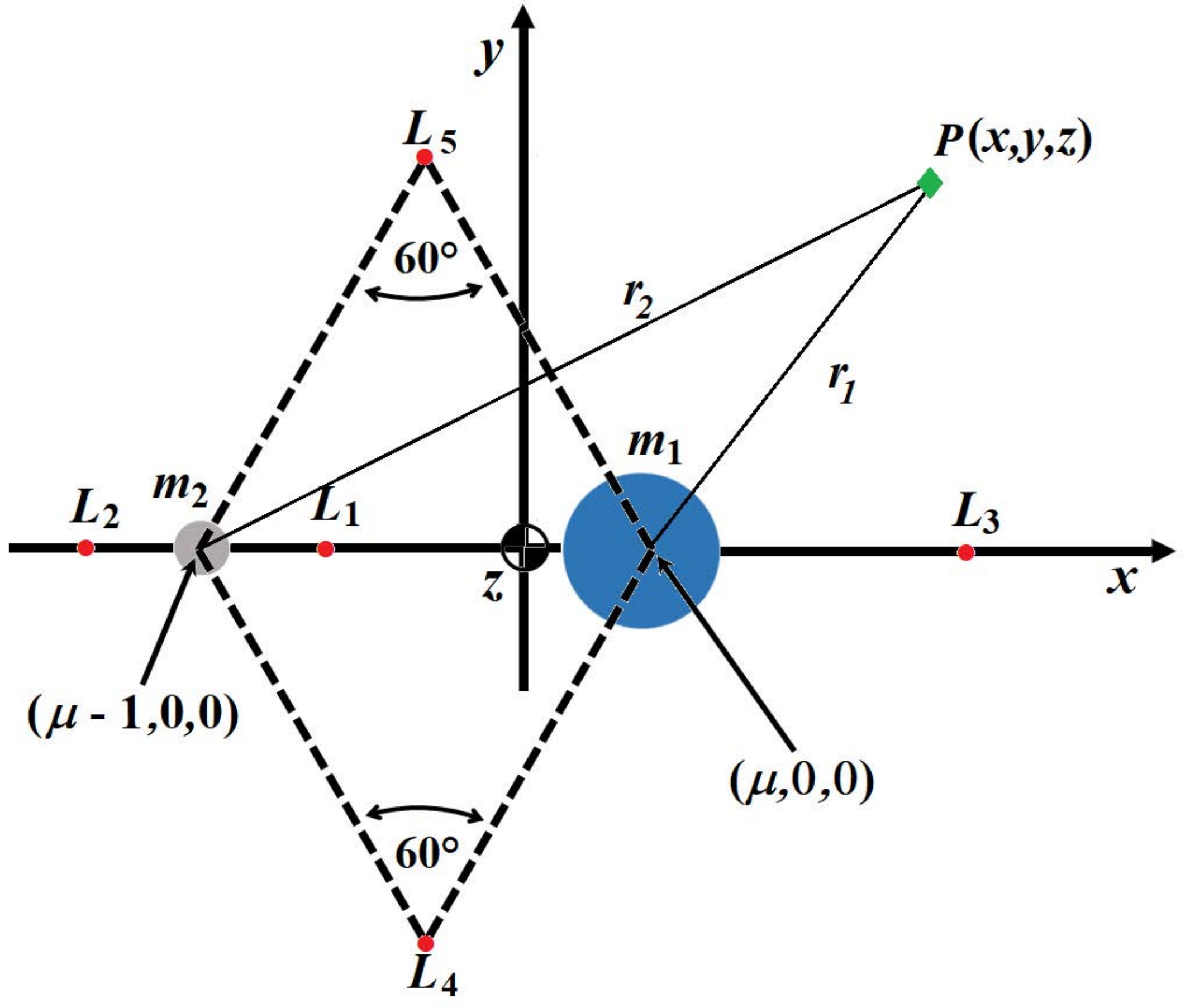}
\caption{Relationship between the barycentric inertial reference frame ($O$, $X$, $Y$, $Z$) and the barycentric synodic reference frame ($O$, $x$, $y$, $z$).}
\label{fig:ref_frames}
\end{figure}

The effect of the mass distribution of the primaries on their mean motion must be included in the model because it enters the equations of motion of the third body. Following the derivation reported in Appendix A,
the acceleration $\ddot{\bf r}_{12}$ of the smaller relative to the larger primary is given by 
\begin{equation}
\ddot{\bf r}_{12} = -\frac{G\left(m_1+m_2\right)}{r^2_{12}}\left[1+\frac{3}{2}\left(\frac{J_2^1 R^2_1 + J_2^2 R^2_2}{r^2_{12}}\right)\right]{\bf u}_{12},
\label{eq:r12}
\end{equation}
where ${\bf r}_{12} = r_{12}{\bf u}_{12}$ denotes the position vector of $m_2$ with respect to $m_1$. Note that
Eq.~\eqref{eq:r12} includes the effects of the quadrupole moment of the mass distribution of each primary over the other considered as a point and neglects the mutual attraction of the two mass distributions.
If $m_1$ and $m_2$ move in circular orbits about their center of mass, 
\begin{equation}
{\bf r}_{12} = \left(\begin{array}{c}
               d\cos{nt} \\ 
							 d\sin{nt} \\
							 0 \\
							\end{array}\right), 
\label{eq:part}
\end{equation}
where $d$ is the constant distance between the primaries and $n$ is their orbital mean motion:
\begin{equation}
n = \sqrt{\frac{G\left(m_1+m_2\right)}{d^3}\left[1 + \frac{3\left(J_2^1R^2_1 + J_2^2R^2_2\right)}{2d^2}\right]}. 
\label{eq:n}
\end{equation}
Normalizing the masses with respect to $m_1+m_2$, the distances with respect to $d$ and the time with respect to a reference value equal to $\sqrt{\frac{d^3}{G(m_1+m_2)}}$ yields unit value for the normalized gravitational constant and the following form for Eq.~\eqref{eq:V}:
\begin{equation}
\hat{V} = \frac{1-\mu}{\hat{r}_1}\left\{1+\frac{A_1}{2\hat{r}^2_1}\left[3\left(\frac{Z}{\hat{r}_1}\right)^2-1\right]\right\} + \frac{\mu}{\hat{r}_2}\left\{1+\frac{A_2}{2\hat{r}^2_2}\left[3\left(\frac{Z}{\hat{r}_2}\right)^2-1\right]\right\}, \label{eq:Vnew}
\end{equation}
where $\mu = m_2/\left(m_1+m_2\right)$, $\hat{r}_1 = r_1/d$, $\hat{r}_2 = r_2/d$,  $A_1 = J_2^1R^2_1/d^2$ and $A_2 = J_2^2 R^2_2/d^2$. In this way, Eq.~\eqref{eq:n} takes the form:
\begin{equation}
\label{eq:nnew}
\hat{n}  =  \sqrt{1+\frac{3\left(A_1+A_2\right)}{2}}.
\end{equation}
Hereinafter, normalization will be assumed unless otherwise specified, but the "hat" symbol will be omitted for the sake of simplicity and clarity.

\subsection{Equations of motion}
The components $\ddot{X}$, $\ddot{Y}$ and $\ddot{Z}$ of the acceleration of the third body in the given inertial frame are
\begin{eqnarray}
\ddot{X} &=& \frac{\partial V}{\partial X}, \label{eq:motionX}\\ 
\ddot{Y} &=& \frac{\partial V}{\partial Y}, \label{eq:motionY}\\ 
\ddot{Z} &=& \frac{\partial V}{\partial Z}. \label{eq:motionZ}
\end{eqnarray}

Let us now consider a reference frame ($O$, $x$, $y$, $z$) centered at $O$, rotating with angular velocity $n$ relative to ($O$, $X$, $Y$, $Z$) and such that the $x$-axis contains the larger primary at $\left(\mu,0,0\right)$ and the smaller primary at $\left(\mu-1,0,0\right)$ and the two frames coincide at $t$ = 0  (Fig.~\ref{fig:ref_frames}). The rotating frame is called barycentric synodic. At any time $t$,
\begin{eqnarray}
X &=& x\cos{nt}-y\sin{nt}, \label{eq:X}\\ 
Y &=& x\sin{nt}+y\cos{nt}, \label{eq:Y}\\ 
Z &=& z. \label{eq:Z}
\end{eqnarray}  
Substituting Eqs.~\eqref{eq:X}-\eqref{eq:Z} and \eqref{eq:Vnew} into Eqs.~\eqref{eq:motionX}-\eqref{eq:motionZ} yields 
\begin{eqnarray}
\ddot{x} - 2n\dot{y} & = & n^2x - \frac{\left(1-\mu\right)\left(x-\mu\right)}{r^3_1}C_1 - \frac{\mu\left(x+1-\mu\right)}{r^3_2}C_2, \label{eq:xsyn} \\ 
\ddot{y} + 2n\dot{x} & = & y\left[n^2 - \frac{\left(1-\mu\right)}{r^3_1}C_1 - \frac{\mu}{r^3_2}C_2\right], \label{eq:ysyn}\\ 
\ddot{z}             & = & -z\left(\frac{1-\mu}{r^3_1}\tilde{C}_1+\frac{\mu}{r^3_2}\tilde{C}_2\right),\label{eq:zsyn}
\end{eqnarray}
in which
\begin{eqnarray}
C_i         & = & 1 - \frac{3}{2}\frac{A_i}{r^2_i}\left[5\left(\frac{z}{r_i}\right)^2-1\right],  \;\;\; i = 1,2 \label{eq:Ci}\\ 
\tilde{C}_i & = & 1 - \frac{3}{2}\frac{A_i}{r^2_i}\left[5\left(\frac{z}{r_i}\right)^2-3\right] = C_i + 3\frac{A_i}{r^2_i},   \;\;\; i = 1,2. \label{eq:Ci_tilde}
\end{eqnarray}
Eqs.~\eqref{eq:xsyn}-\eqref{eq:zsyn} are the equations of motion of the $J_2$-perturbed CR3BP in the synodic barycentric reference frame. Then, introducing the effective potential $\Omega$
\begin{eqnarray}
\Omega & = &  
\left(1-\mu\right)\left\{\frac{1}{r_1}+\frac{3}{2}\frac{A_1}{r^3_1}\left[\frac{1}{3}-\left(\frac{z}{r_1}\right)^2\right]\right\} 
+ \mu\left\{\frac{1}{r_2}+\frac{3}{2}\frac{A_2}{r^3_2}\left[\frac{1}{3}-\left(\frac{z}{r_2}\right)^2\right]\right\}  \nonumber \\ 
& +& \frac{n^2}{2}\left[\left(1-\mu\right)r^2_1+\mu r^2_2-z^2\right] + \frac{\mu(1-\mu)}{2} \label{eq:Omega}
\end{eqnarray}
allows to rewrite Eqs.~\eqref{eq:xsyn}-\eqref{eq:zsyn} as
\begin{eqnarray}
\ddot{x} - 2n\dot{y} & = & \displaystyle \Omega_x, \label{eq:xOmega}\\ 
\ddot{y} + 2n\dot{x} & = & \displaystyle \Omega_y, \label{eq:yOmega}\\ 
\ddot{z}             & = & \displaystyle \Omega_z, \label{eq:zOmega}
\end{eqnarray}
in which the notation $\Omega_x$ denotes partial differentiation of $\Omega$ with respect to $x$ and similarly for the other two components.
The quantity $C_J$ defined as
\begin{equation}
C_J = 2\Omega - \left(\dot{x}^2+\dot{y}^2+\dot{z}^2\right) \label{eq:CJ}
\end{equation}
is a constant of motion. It is the equivalent of the Jacobi constant in this model.

\subsection{Equilibrium points}
Similarly to the unperturbed CR3BP, this model admits five equilibrium positions in the $xy$-plane, three of which (the collinear equilibria) lie on the $x$-axis. 
Their $x$-coordinates $x_{L_i}$  ($i = 1, 2, 3$) are obtained by imposing $\Omega_x$ = $\Omega_y$ = $\Omega_z$ = 0 and $y$ = $z$ = $0$ in
Eqs.~\eqref{eq:xsyn}-\eqref{eq:zsyn} and solving the resulting quintic polynomial in $x$
\begin{eqnarray}
& & x\left[1+\frac{3}{2}\left(A_1+A_2\right)\right] - \left(1-\mu\right)\left(x-\mu\right)\left(\frac{1}{\left|x-\mu\right|^3}+\frac{3A_1}{2\left|x-\mu\right|^5}\right) \nonumber\\
& - & \mu\left(x-\mu+1\right)\left(\frac{1}{\left|x-\mu+1\right|^3}+\frac{3A_2}{2\left|x-\mu+1\right|^5}\right) = 0 \label{eq:quintic}
\end{eqnarray}
for the three cases identified by the possible combinations of signs of the expressions in absolute value. 
Such solutions can be approximated, for example, with a Newton-Raphson scheme. 

The coordinates  $x_{L_4}$, $y_{L_4}$, $x_{L_5}$ and $y_{L_5}$  of the triangular points $L_4$ and $L_5$ follow from $\Omega_x$ = $\Omega_y$ =  $\Omega_z$ = 0, $z$ = 0 and $y\neq 0$. The results have the following analytical expressions:
\begin{eqnarray}
x_{L_4,L_5} & = & \mu - \frac{1}{2} -\frac{1}{2}\left(A_1-A_2\right) + \frac{5}{8}\left(A^2_1-A^2_2\right),\label{eq:x45}\\
y_{L_4,L_5} & = & \pm\frac{\sqrt{3}}{2}\left\{1-\frac{1}{3}\left(A_1-A_2\right)+\frac{1}{36}\left[7\left(A^2_1+A^2_2\right)+68A_1A_2\right]\right\},\label{eq:y45}
\end{eqnarray}
where $+$ and $-$ apply to $L_4$ and $L_5$, respectively.

\subsection{The state transition matrix}
In the vicinity of equilibrium points and LPOs of the CR3BP, the nonlinear equations of motion of the system can be linearized by a Taylor series expansion which expresses the variation in the state $\delta {\bf s}$ as a first-order differential form of a linear system (see, e.g., \cite{Wiesel:2003}):
\begin{equation}
\label{eq:var_state}
\delta \dot{\bf s} = A(t)  \delta {\bf s},
\end{equation}
with $A(t)$ a time-varying matrix.
The above can be extended to the $J_2$-perturbed dynamical model described by Eqs.~\eqref{eq:xsyn}-\eqref{eq:zsyn}.  In this case, 
$A(t)$ is given by
\begin{equation}
A\left(t\right)=
\left[
\begin{array}{c|c}
\boldsymbol{O} & \boldsymbol{I} \\
\hline
\boldsymbol{U} & \boldsymbol{J}
\end{array}
\right]. \label{eq:A}
\end{equation}
$\boldsymbol {O}$ is the $3\times 3$ null matrix, $\boldsymbol {I}$ is the $3\times3$ identity matrix, whereas $\boldsymbol{J}$ and $\boldsymbol{U}$ are defined as 
\begin{eqnarray}
\boldsymbol{J} & = & 
\left[
\begin{array}{c c c}
0 & 2n & 0\\
-2n & 0 & 0\\
0 & 0 & 0
\end{array}
\right],\label{eq:J} \\ 
\boldsymbol{U} & = &
\left[
\begin{array}{c c c}
\Omega_{xx} & \Omega_{xy} & \Omega_{xz}\\
\Omega_{xy} & \Omega_{yy} & \Omega_{yz}\\
\Omega_{xz} & \Omega_{yz} & \Omega_{zz}
\end{array}
\right]. \label{eq:U}
\end{eqnarray}
The elements of $\boldsymbol{U}$ have the following expressions: 
\begin{eqnarray}
\Omega_{xx} & = & n^2 - \frac{\left(1-\mu\right)}{r^3_1}C_1-\frac{\mu}{r^3_2}C_2\nonumber\\
            & + & \left(x-\mu\right)^2\left\{\frac{3\left(1-\mu\right)}{r^5_1}C_1 - \frac{3\left(1-\mu\right)}{r^7_1}A_1\left[10\left(\frac{z}{r_1}\right)^2-1\right]\right\}\nonumber\\
            & + & \left(x+1-\mu\right)^2\left\{\frac{3\mu}{r^5_2}C_2 - \frac{3\mu}{r^7_2}A_2\left[10\left(\frac{z}{r_2}\right)^2-1\right]\right\}, \label{eq:Omegaxx} 
\end{eqnarray}
\begin{eqnarray}
\Omega_{xy} & = & y\left\{\frac{3\left(1-\mu\right)\left(x-\mu\right)}{r^5_1}C_1 - \frac{3\left(1-\mu\right)\left(x-\mu\right)}{r^7_1}A_1\left[10\left(\frac{z}{r_1}\right)^2-1\right]\right\}\nonumber\\
            & + & y\left\{\frac{3\mu\left(x+1-\mu\right)}{r^5_2}C_2 - \frac{3\mu\left(x+1-\mu\right)}{r^7_2}A_2\left[10\left(\frac{z}{r_2}\right)^2-1\right]\right\} , \label{eq:Omegaxy} 
\end{eqnarray}
\begin{eqnarray}
\Omega_{xz} & = & z\left\{\frac{3\left(1-\mu\right)\left(x-\mu\right)}{r^5_1}C_1 - \frac{3\left(1-\mu\right)\left(x-\mu\right)}{r^7_1}A_1\left[10\left(\frac{z}{r_1}\right)^2-6\right]\right\}\nonumber\\
            & + & z\left\{\frac{3\mu\left(x+1-\mu\right)}{r^5_2}C_2 - \frac{3\mu\left(x+1-\mu\right)}{r^7_2}A_2\left[10\left(\frac{z}{r_2}\right)^2-6\right]\right\}, \label{eq:Omegaxz} 
\end{eqnarray}
\begin{eqnarray}
\Omega_{yy} & = & n^2 - \frac{\left(1-\mu\right)}{r^3_1}C_1-\frac{\mu}{r^3_2}C_2\nonumber\\
            & + & y^2\left\{\frac{3\left(1-\mu\right)}{r^5_1}C_1 - \frac{3\left(1-\mu\right)}{r^7_1}A_1\left[10\left(\frac{z}{r_1}\right)^2-1\right]\right\}\nonumber\\
            & + & y^2\left\{\frac{3\mu}{r^5_2}C_2 - \frac{3\mu}{r^7_2}A_2\left[10\left(\frac{z}{r_2}\right)^2-1\right]\right\}, \label{eq:Omegayy} 
\end{eqnarray}
\begin{eqnarray}
\Omega_{yz} & = & zy\left\{\frac{3\left(1-\mu\right)}{r^5_1}C_1 - \frac{3\left(1-\mu\right)}{r^7_1}A_1\left[10\left(\frac{z}{r_1}\right)^2-6\right]\right\}\nonumber\\
            & + & zy\left\{\frac{3\mu }{r^5_2}C_2 - \frac{3\mu }{r^7_2}A_2\left[10\left(\frac{z}{r_2}\right)^2-6\right]\right\}, \label{eq:Omegayz} 
\end{eqnarray}
\begin{eqnarray}
\Omega_{zz} & = & -\frac{\left(1-\mu\right)}{r^3_1}\tilde{C}_1-\frac{\mu}{r^3_2}\tilde{C}_2\nonumber\\
            & + & z^2\left\{\frac{3\left(1-\mu\right)}{r^5_1}\tilde{C}_1 - \frac{3\left(1-\mu\right)}{r^7_1}A_1\left[10\left(\frac{z}{r_1}\right)^2-8\right]\right\}\nonumber\\
            & + & z^2\left\{\frac{3\mu}{r^5_2}\tilde{C}_2 - \frac{3\mu}{r^7_2}A_2\left[10\left(\frac{z}{r_2}\right)^2-8\right]\right\}. \label{eq:Omegazz}
\end{eqnarray}
The general solution to Eq.~\eqref{eq:var_state} is 
\begin{equation}
\label{eq:var_sol}
\delta {\bf s}(t) = \Phi\left(t,t_0\right)  \delta {\bf s}_0,
\end{equation}
where $\delta {\bf s}_0$ is the initial condition for the variation and $\Phi\left(t,t_0\right) = \displaystyle \frac{\partial {\bf s}(t)}{\partial {\bf s}_0}$
is the state transition matrix. $\Phi$ satisfies the differential equation
\begin{equation}
\dot{\Phi}\left(t,t_0\right) = A\left(t\right)\Phi\left(t,t_0\right). \label{eq:Phi}
\end{equation}
The elements of $\Phi$ are required for any differential correction process based on a Newton-Raphson iteration scheme and for the approximation of the initial state of the HIMs of LPOs.

\section{Application to the Saturn-Enceladus system}
\label{sec:app}
The application of the above model to the system composed by Saturn, Enceladus and the spacecraft is made possible by the fact that the orbit of Enceladus is approximately contained in the equatorial plane of Saturn and its axial tilt is negligible. The parameters of the system including the oblateness of the two primaries are reported in Table~\ref{tab:SE}. Hereinafter, all quantities referred to Saturn will be labeled with $S$ instead of $1$, whereas for Enceladus $E$ will replace the index $2$.
We distinguish among three $J_2$-perturbed CR3BPs: the model perturbed by $J^{S}_2$ only (CR3BP + $J^{S}_2$), that perturbed by $J^{E}_2$ only (CR3BP + $J^{E}_2$) and the full model (CR3BP + $J^{S}_2$ + $J^{E}_2$). This is done in order to gain insight into the relative importance of the two oblateness effects.

Tables~\ref{tab:L123} and \ref{tab:L45} report the positions of the equilibrium points for the unperturbed CR3BP and for the three perturbed models. Tables~\ref{tab:varL123} and \ref{tab:varL45} present the displacements of the equilibrium points in each perturbed model with respect to the unperturbed CR3BP.  
\begin{table}[h!]
\caption{Parameters of the $J_2$-perturbed Saturn-Enceladus CR3BP: equatorial radius $R_S$ of Saturn, equatorial radius $R_E$ of Enceladus, mass ratio $\mu$, distance $d$ between Saturn and Enceladus, second zonal harmonic coefficient $J^{S}_2$ of Saturn, second zonal harmonic coefficient $J^{E}_2$ of Saturn \cite{Horizon, Campbell:1989}.}
\label{tab:SE}
\begin{center}
\begin{tabular}{cccccc}
 \hline
$R_{S}$ & $R_{E}$ & $\mu$ & $d$ & $J^{S}_2$ & $J^{E}_2$ \\
(km) & (km) & ($\cdot 10^{-6}$) & (km) & & \\
\hline
60268.0 & 252.1 &$ {\bf 0.190}$ & 238042.0 & $1.6298\cdot10^{-2}$ &$2.5\cdot 10^{-3}$\\
\hline
\end{tabular}
\end{center}
\end{table}
\begin{table}[h!]
\caption{Synodic barycentric $x$-coordinates of the collinear equilibrium points in the Saturn-Enceladus CR3BP and in each of the three $J_2$-perturbed models.}
\label{tab:L123}
\begin{center}
\begin{tabular}{lccc}
 \hline
Model  & $x_{L_1}$ & $x_{L_2}$ & $x_{L_3}$ \\ 
\hline
CR3BP &             -0.9960195     & -1.0039907      & 1.0000001 \\ 
CR3BP + $J^{E}_2$ & -0.9960192     & -1.0039910      & 1.0000001 \\ 
CR3BP + $J^{S}_2$ & -0.9960230     & -1.0039872      & 1.0000001 \\ 
CR3BP + $J^{E}_2$ + $J^{S}_2$ & -0.9960226  & -1.0039876 & 1.0000001\\ 
 \hline
\end{tabular}
\end{center}
\end{table}

\begin{table}[h!]
\caption{Synodical barycentric $x$- and $y$-coordinates of the triangular points in the Saturn-Enceladus CR3BP and in each of the three $J_2$-perturbed models.}
\label{tab:L45}
\begin{center}
\begin{tabular}{lcc}
 \hline
Model      & $x_{L_4,L_5}$ & $y_{L_4,L_5}$ \\ 
\hline
CR3BP             & -0.4999998      & $\pm$ 0.8660254 \\ 
CR3BP + $J^{E}_2$ & -0.4999998      & $\pm$ 0.8660254  \\ 
CR3BP + $J^{S}_2$ & -0.5005215      & $\pm$ 0.8657240  \\ 
CR3BP + $J^{E}_2$ + $J^{S}_2$ & -0.5005215  & $\pm$ 0.8657240 \\ 
 \hline
\end{tabular}
\end{center}
\end{table}

\begin{table}[h!]
\caption{Displacements $\Delta x_{L_i}$ ($i$ = 1,2,3) of the $x$-coordinates of the collinear equilibrium points in the $J_2$-perturbed Saturn-Enceladus CR3BPs with respect to the unperturbed model.}
\label{tab:varL123}
\begin{center}
\begin{tabular}{lrrr}
 \hline
Model & $\Delta x_{L_1}$ & $\Delta x_{L_2}$ & $\Delta x_{L_3}$ \\
&  (km)  & (km) & (km) \\
 \hline
CR3BP + $J^{E}_2$ & 0.1 & 0.1 & $\sim$0 \\ 
CR3BP + $J^{S}_2$ & 0.8 & 0.8 & $\sim$0\\ 
CR3BP + $J^{E}_2$ + $J^{S}_2$ & 0.7 & 0.7 & $\sim$0\\ 
 \hline
\end{tabular}
\end{center}
\end{table}

\begin{table}[h!]
\caption{Displacements $\Delta x_{L_i}$ and $\Delta y_{L_i}$ ($i$ = 4.5)  of the $x$- and $y$-coordinates of the triangular equilibrium points in the $J_2$-perturbed Saturn-Enceladus CR3BPs with respect to the unperturbed model.}
\label{tab:varL45}
\begin{center}
\begin{tabular}{lrr}
 \hline
Model  & $\Delta x_{L_4,L_5}$ & $\Delta y_{L_4,L_5}$ \\ 
  &  (km)  & (km)  \\
 \hline
CR3BP + $J^{E}_2$ & $\sim$0 & $\sim$0  \\
CR3BP + $J^{S}_2$ & 124.2 & 71.7  \\  
CR3BP + $J^{E}_2$ + $J^{S}_2$ & 124.2  & 71.7 \\  \hline
\end{tabular}
\end{center}
\end{table}

\subsection{Halo orbits}
\begin{figure}[h!]
	\centering
	\includegraphics[scale = 0.175]{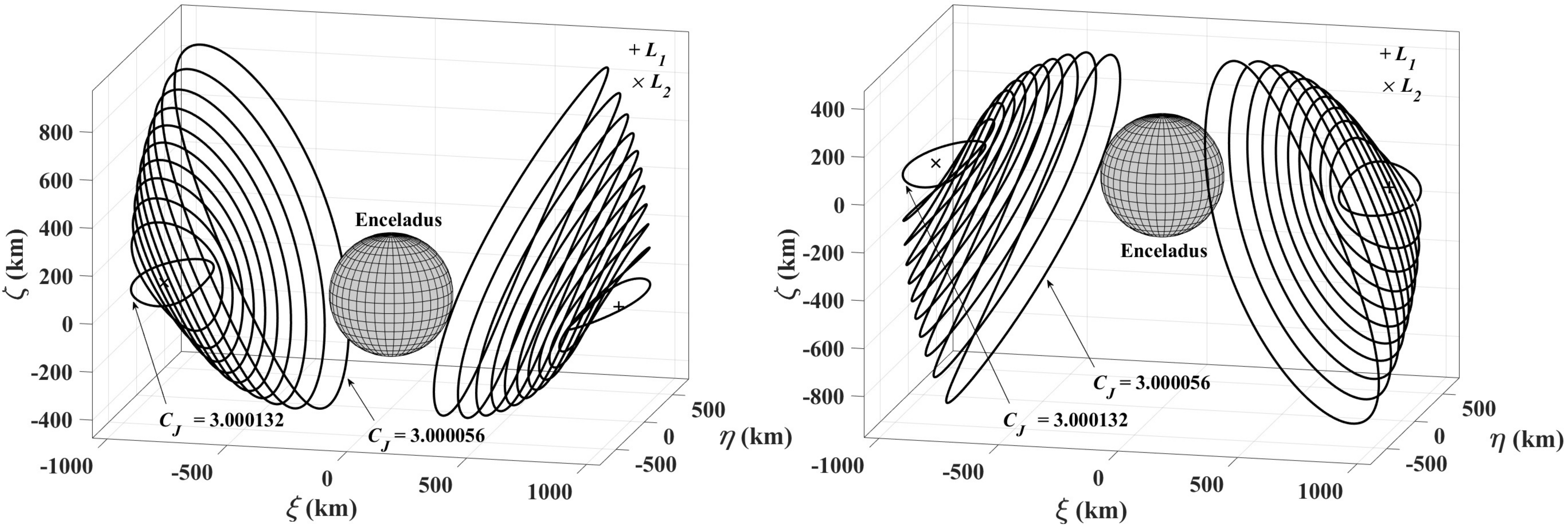}
	\caption{Families of Northern (left) and Southern (right) halo orbits around $L_1$ and $L_2$ in the Saturn-Enceladus CR3BP (Enceladus-centered synodical reference frame) (from \cite{Fantino:2020}).}
	\label{fig:HalosCR3BP}
\end{figure}

The linear approximation of the dynamics of the CR3BP in the neighborhood of a given equilibrium point leads to families of LPOs \cite{Szebehely:1967}. \cite{Davis:2018} and \cite{Fantino:2020} presented families of Northern and Southern halo orbits around $L_1$ and $L_2$ in the Saturn-Enceladus CR3BP. In particular, \cite{Fantino:2020} explored these orbits in a wide Jacobi constant range, from $3.000055$ to $3.000131$ (see Fig.~\ref{fig:HalosCR3BP} where the $\xi$, $\eta$ and $\zeta$ axes are parallel to $x$, $y$ and $z$ and the origin is at the center of Enceladus). 
Note that a Southern halo orbit can be obtained from a Northern halo orbit through the transformation $z \rightarrow -z,$ $\dot{z} \rightarrow -\dot{z}$.    

\begin{figure}[h!]
	\centering
	\includegraphics[scale = 0.2]{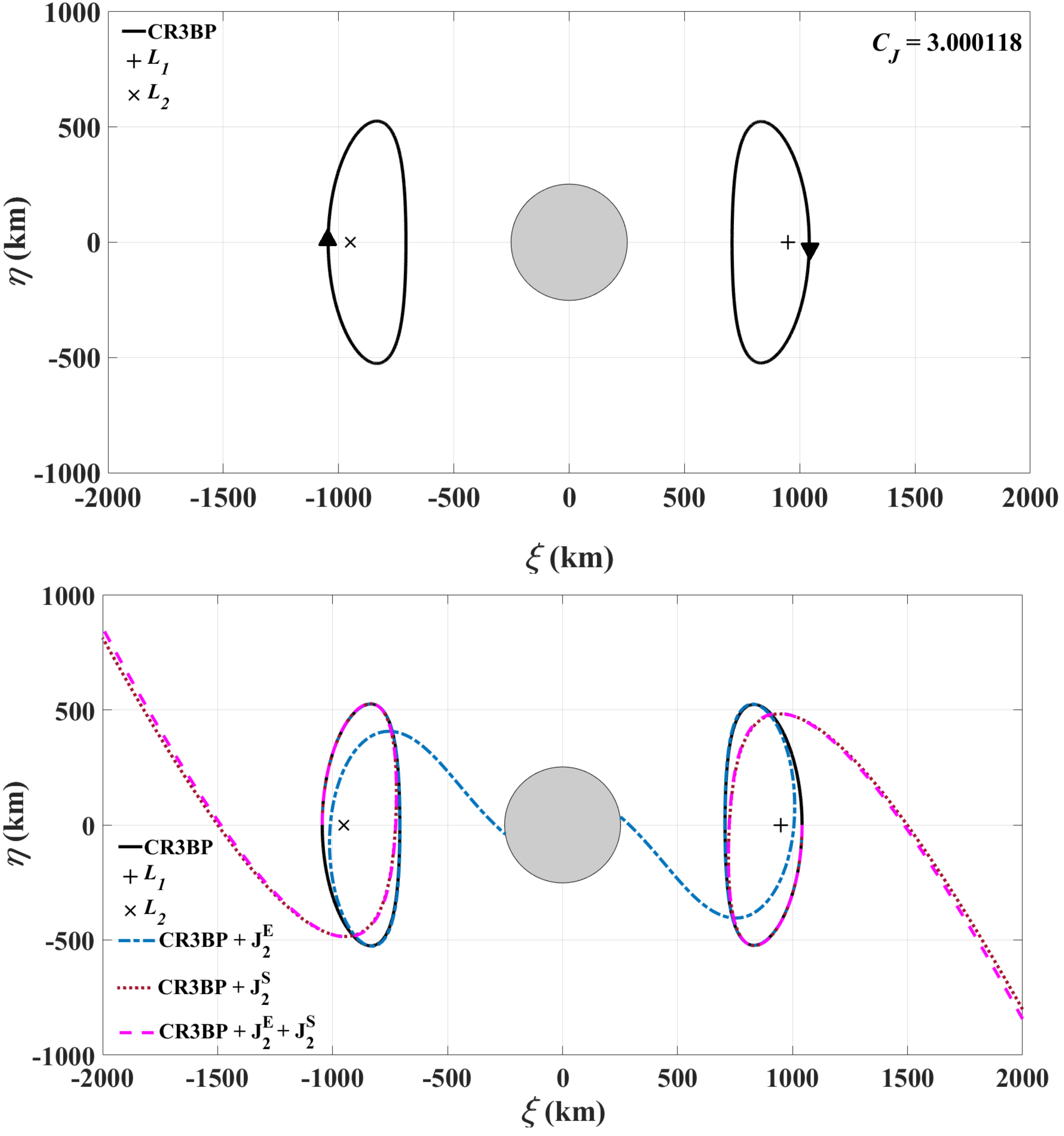}
	\caption{Top: planar projection of two halo orbits with $C_J = 3.000118$ in the Saturn-Enceladus CR3BP. Bottom: effects of the perturbations of $J^{E}_2$ and $J^{S}_2$ individually and in combination. Enceladus-centered synodical frame.}
	\label{fig:Halo_dev}  
\end{figure}

The oblateness of Saturn and Enceladus perturb these solutions significantly, as illustrated in Fig.~\ref{fig:Halo_dev} which shows the projection on the plane of the primaries of the evolution of initial conditions corresponding to two Northern halo orbits around $L_1$ and $L_2$ with $C_J = 3.000118$  when $J^{E}_2$ and $J^{S}_2$ are brought into the dynamics: starting from the point represented with triangle markers in the top panel, the oblateness of the primaries causes deviations from the unperturbed periodic orbit which may lead to escapes from the system or collisions with the moon. 

\begin{figure}[h!]
	\centering
	\includegraphics[scale = 0.2]{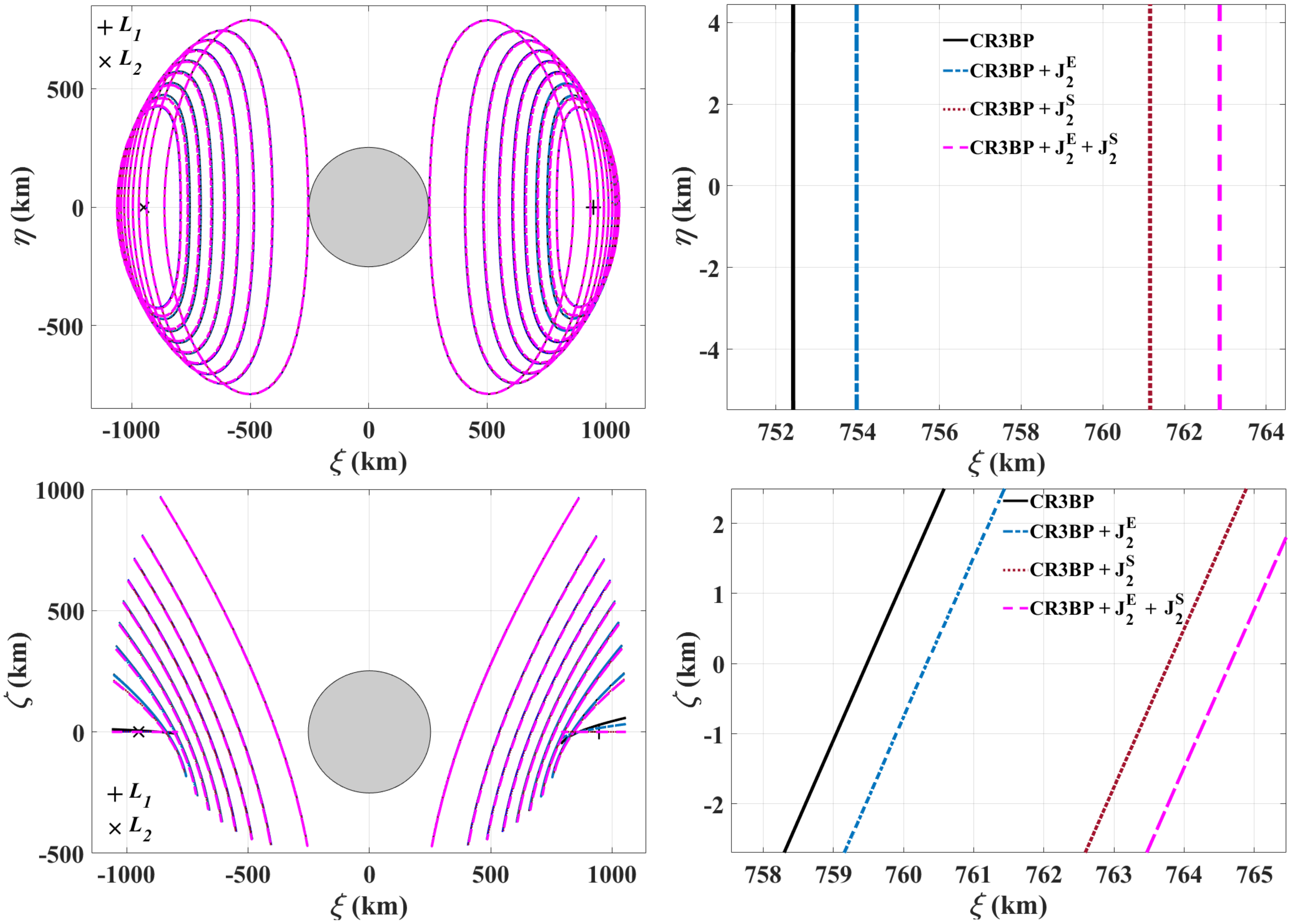}
	\caption{Left: planar projections of halo orbits about $L_1$ and $L_2$ in the Saturn-Enceladus system in the unperturbed and perturbed models. Right: zoomed views of the left plots. Enceladus-centered synodical reference frame.}
	\label{fig:Halo_pert}
\end{figure}

However, dynamical substitutes of the unperturbed halo orbits, i.e., periodic orbits obtained by refining in the new model initial states corresponding to halo orbits in the CR3BP, do exist in the $J_2$-perturbed models. They have been obtained from initial conditions of the unperturbed problem by differential correction imposing $xz$-plane symmetry and periodicity.
Since Eqs.~\eqref{eq:xsyn}-\eqref{eq:zsyn} are invariant under the transformations $y\rightarrow-y$ and $t\rightarrow-t$, the numerical algorithm presented by \cite{Howell:1984} for the classical CR3BP can be extended to the perturbed models. Here, the initial guess is a state on the crossing of an unperturbed halo orbit with the $xz$-plane. At the next crossing of the  same plane, the propagated state and state transition matrix are employed to correct the initial state in order to meet the symmetry requirement and eventually obtain a periodic solution in the perturbed model. This method has been applied to calculate families of halo orbits around $L_1$ and $L_2$ uniformly distributed in $C_J$ and accounting for the oblateness of Saturn and Enceladus separately (CR3BP + $J^{E}_2$, CR3BP +  $J^{S}_2$) and in combination (CR3BP + $J^{E}_2$ + $J^{S}_2$). 
Figure~\ref{fig:Halo_pert} illustrates planar projections of the two families of Northern halo orbits about $L_1$ and $L_2$ in the unperturbed CR3BP and in the  three perturbed models. The Jacobi constant ranges are [3.000055, 3.000132] for CR3BP + $J^{E}_2$ and [3.002668, 3.002744] for CR3BP + $J^{S}_2$ and CR3BP + $J^{E}_2$ + $J^{S}_2$. The right panels of the figure zoom into portions of the left plots to show the deviations of the perturbed solutions from the unperturbed ones.

\subsection{J2-perturbed heteroclinic connections between halo orbits}
\begin{figure}[h!]
\centering
\includegraphics[scale = 0.20]{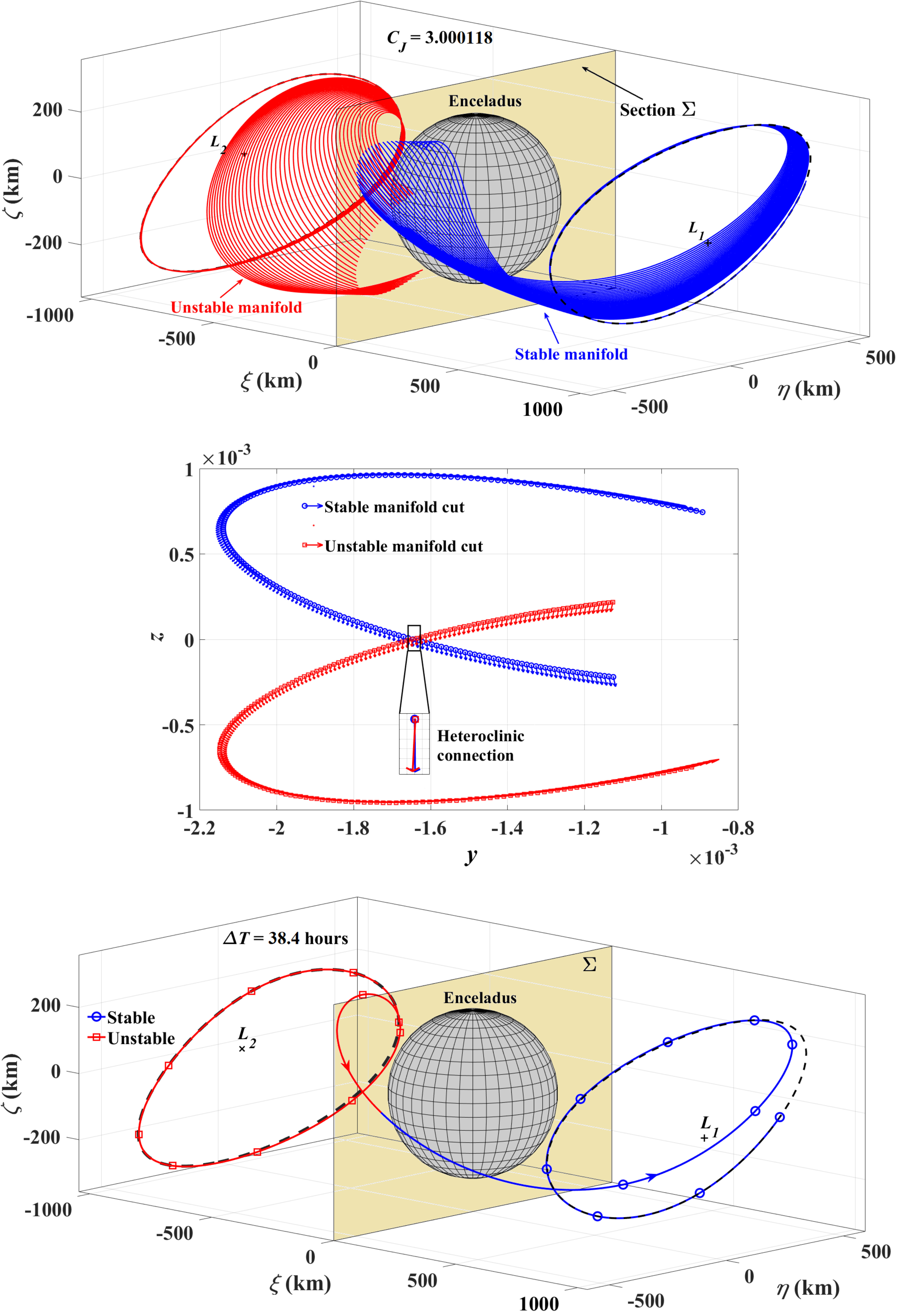}
\caption{Top: non-impacting trajectories of the unstable (red) and stable (blue) HIMs respectively from a Southern $L_2$  halo and a Northern $L_1$ halo with $C_J = 3.000118$. Middle: vector representation of the intersections with $\Sigma$. Bottom: 3D view of the heteroclinic connection (Enceladus-centered synodic reference frame) (from \cite{Fantino:2020}).}
\label{fig:Hetero_Method}  
\end{figure}

Heteroclinic connections between $J_2$-perturbed halo orbits have been computed in the three models.
Stable and unstable hyperbolic invariant manifolds of two orbits with the same Jacobi constant have been computed (the size of the displacement from the halo orbit to the first state on the manifold  is $10^{-6}$ in normalized units) and propagated up to a surface of section $\Sigma$ defined by $x = -1+\mu$. A minimum allowed altitude of 20 km from the surface of Enceladus is set as a safety measure against collisions with the moon. Furthermore, given the symmetries of the problem, only crossings with $\dot{x} > 0$ forward in time are considered.
For every trajectory, the intersection with $\Sigma$ is represented by four state components (i.e., $x$, $y$, $\dot{y}$, $\dot{z}$), whereas $\dot{x}$ is determined by the given $C_J$. These components can be displayed using a vectorial representation of position $\left(y,z\right)$ and velocity $\left(\dot{y},\dot{z}\right)$. In this work, finding a connection means identifying the pair of unstable and stable orbits with minimum position and velocity difference (see Fig.~\ref{fig:Hetero_Method}). Alternative methods have been adopted in other contributions. For example,  \cite{Haapala:2014} uses a single segment to represent simultaneously four states: two states are indicated by the coordinates of the segment base-point, and two additional coordinates are represented by the length. \cite{Geisel:2013} represents $y$, $z$ and $\dot{y}$ in a three-dimensional visual environment in which $\dot{z}$ is displayed using color. \cite{Paskowitz:2006a} chooses spherical coordinates to represent the states at the closest approach to the primary (periapsis map).

\begin{figure}[h!]
\centering
\includegraphics[scale = 0.15]{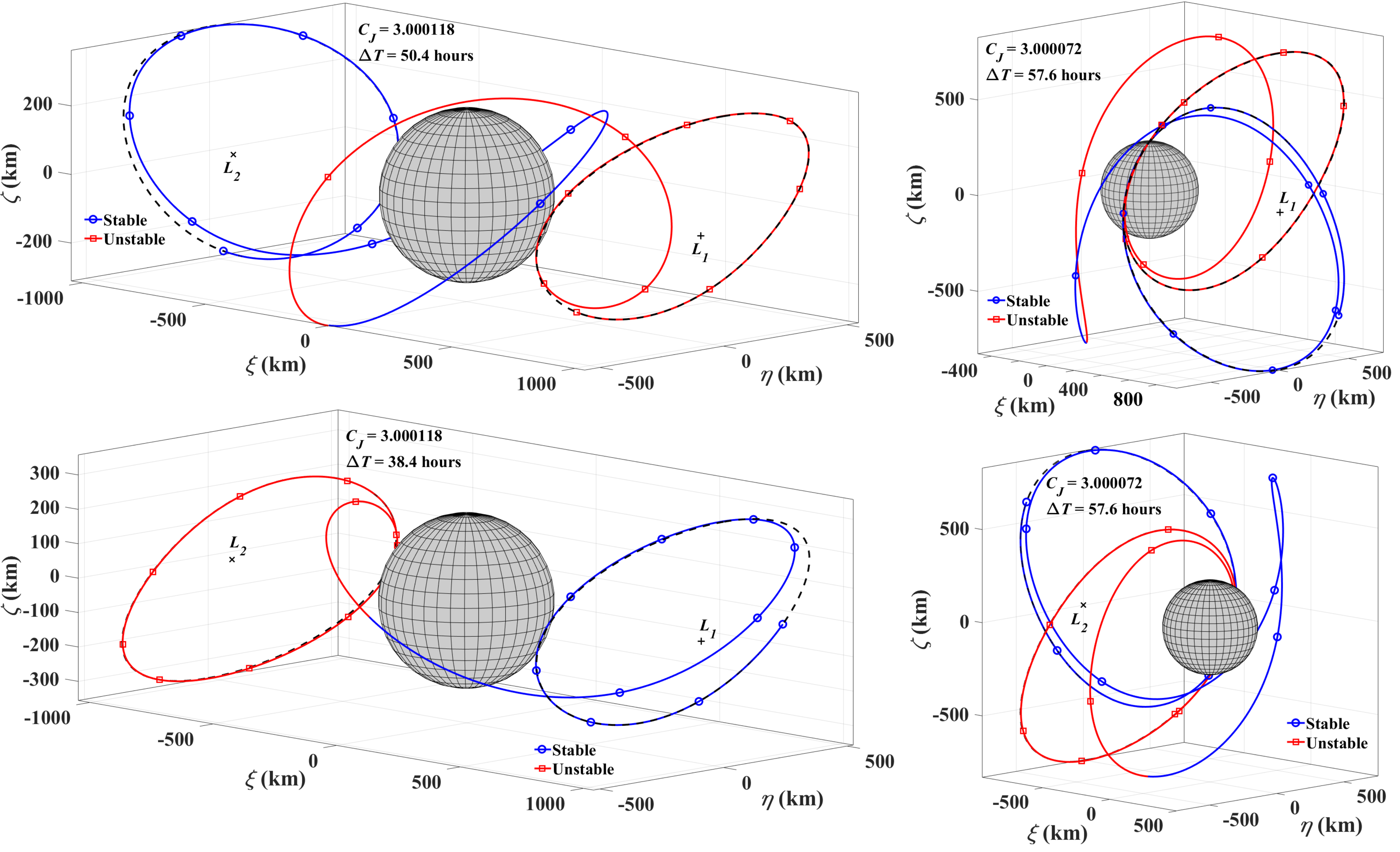}
\caption{Heteroclinic connections of type A (top left), B (bottom left), C (top right), and D (bottom right) in the Saturn-Enceladus unperturbed CR3BP (Enceladus-centered synodic reference frame).}
\label{fig:Heteros_CR3BP}  
\end{figure}

\begin{figure}[h!]
\centering
\includegraphics[scale = 0.22]{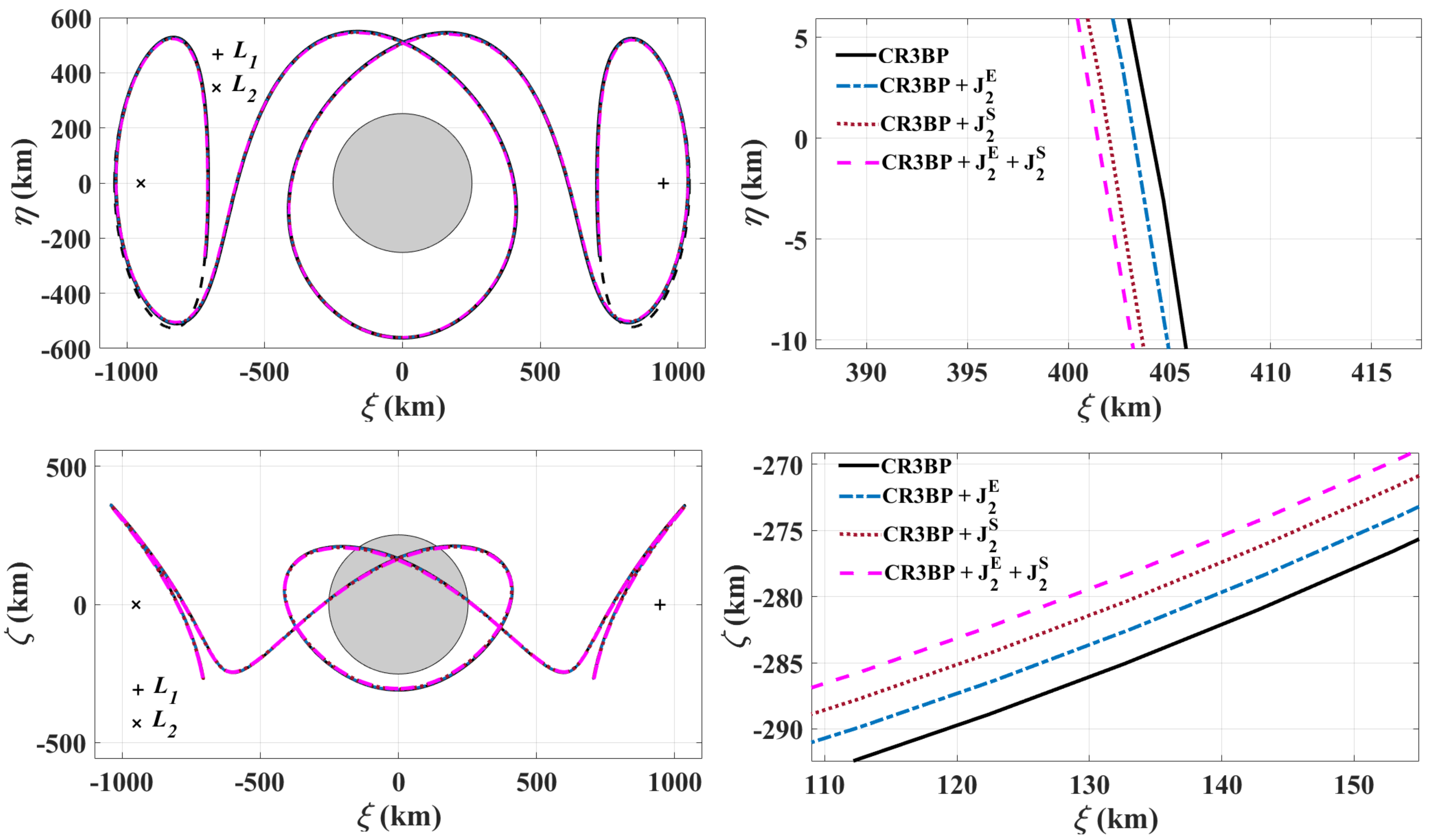}
\caption{Left: planar projections of the heteroclinic connection type A in the three $J_2$-perturbed Saturn-Enceladus CR3BPs (Enceladus-centered synodical frame). Right: zoomed views of the left plots.}
\label{fig:TypeA_Pert}
\end{figure}

\begin{figure}[h!]
\centering
\includegraphics[scale = 0.22]{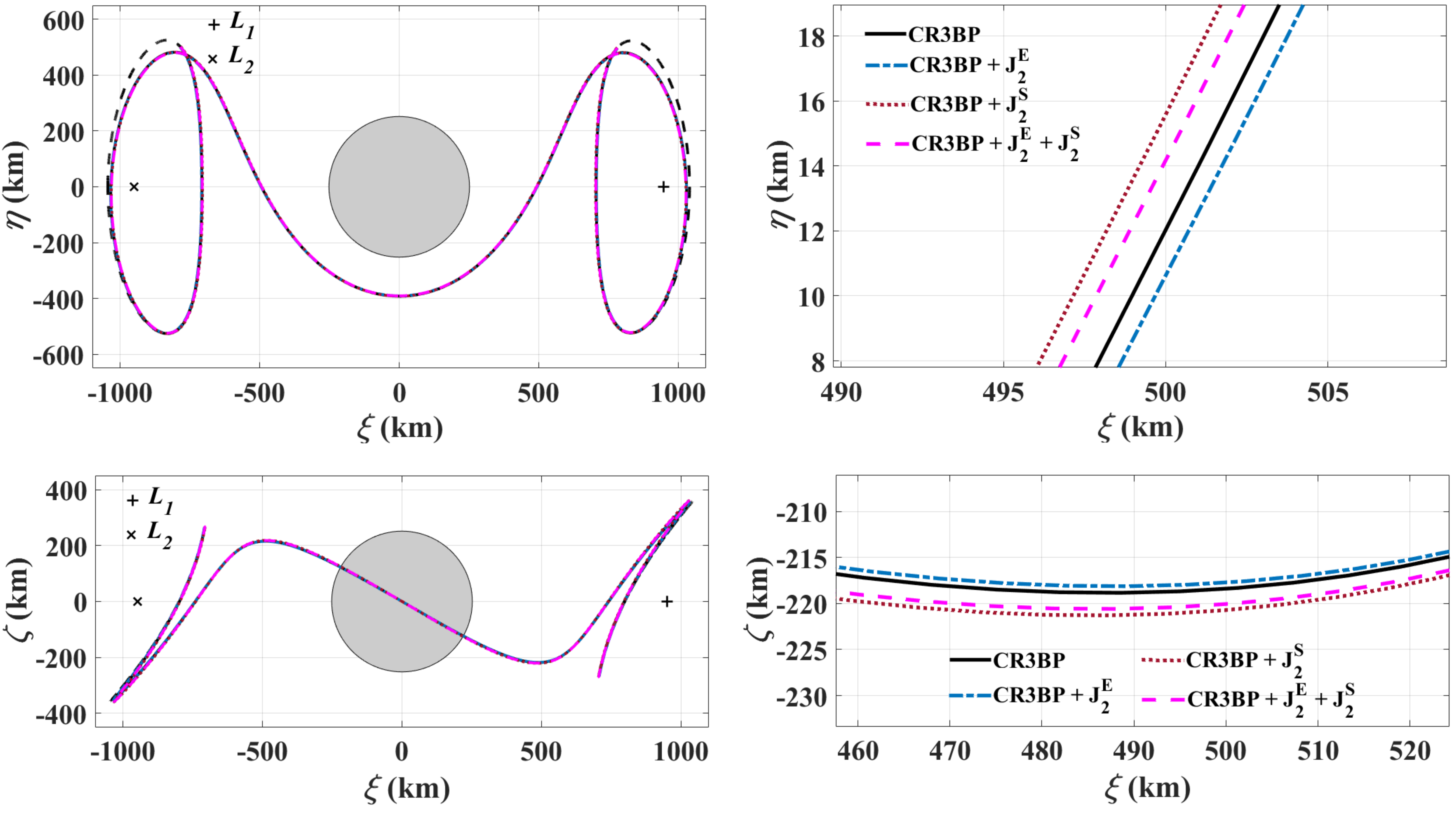}	
\caption{Left: planar projections of the heteroclinic connection type B in the three $J_2$-perturbed Saturn-Enceladus CR3BPs (Enceladus-centered synodical frame). Right: zoomed views of the left plots.}
\label{fig:TypeB_Pert}
\end{figure}

\begin{figure}[h!]
\centering
\includegraphics[scale = 0.22]{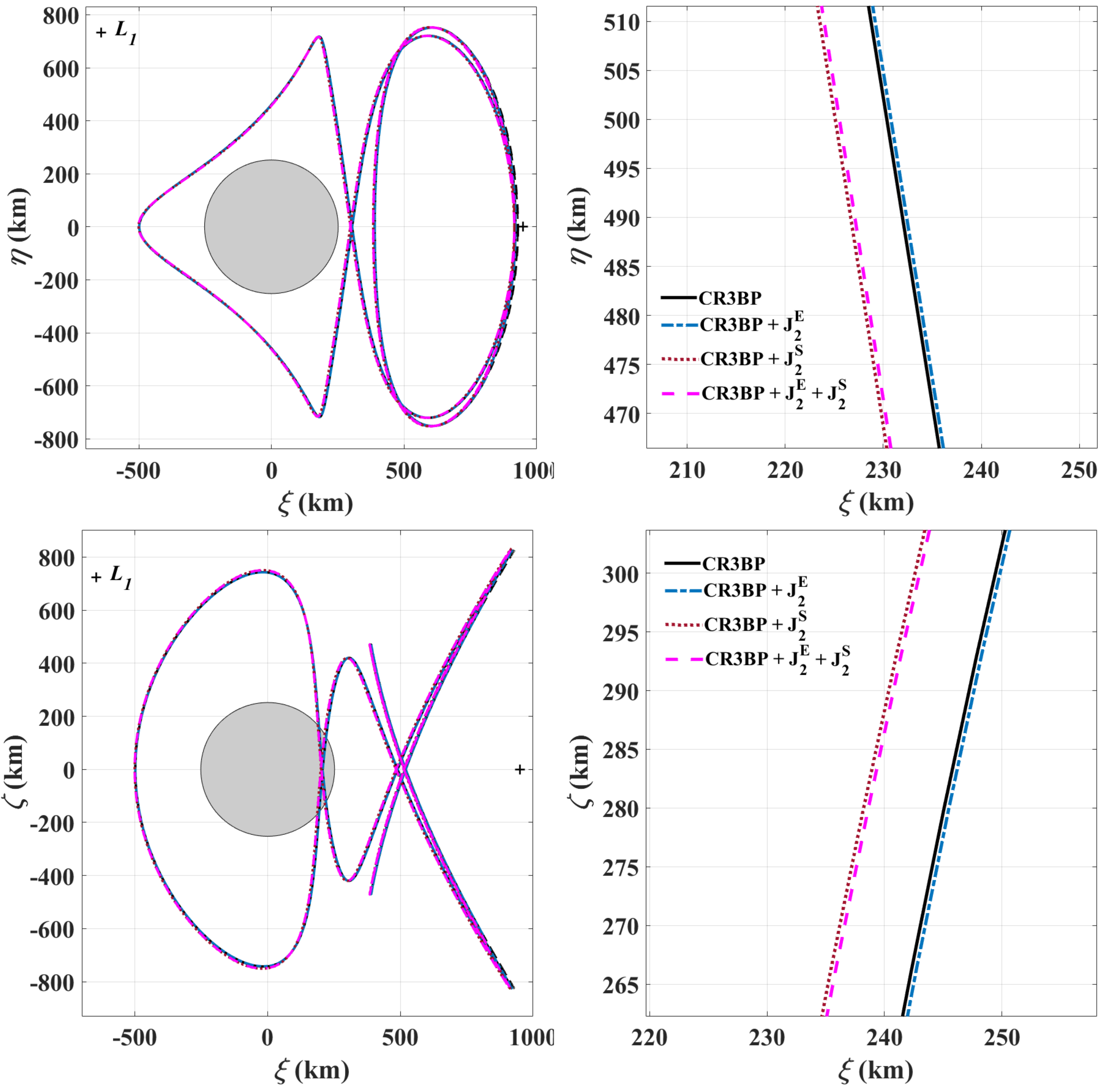}
\caption{Left: planar projections of the heteroclinic connection type C in the three $J_2$-perturbed Saturn-Enceladus CR3BPs (Enceladus-centered synodical frame). Right: zoomed views of the left plots.}
\label{fig:TypeC_Pert}
\end{figure}

\begin{figure}[h!]
\centering
\includegraphics[scale = 0.22]{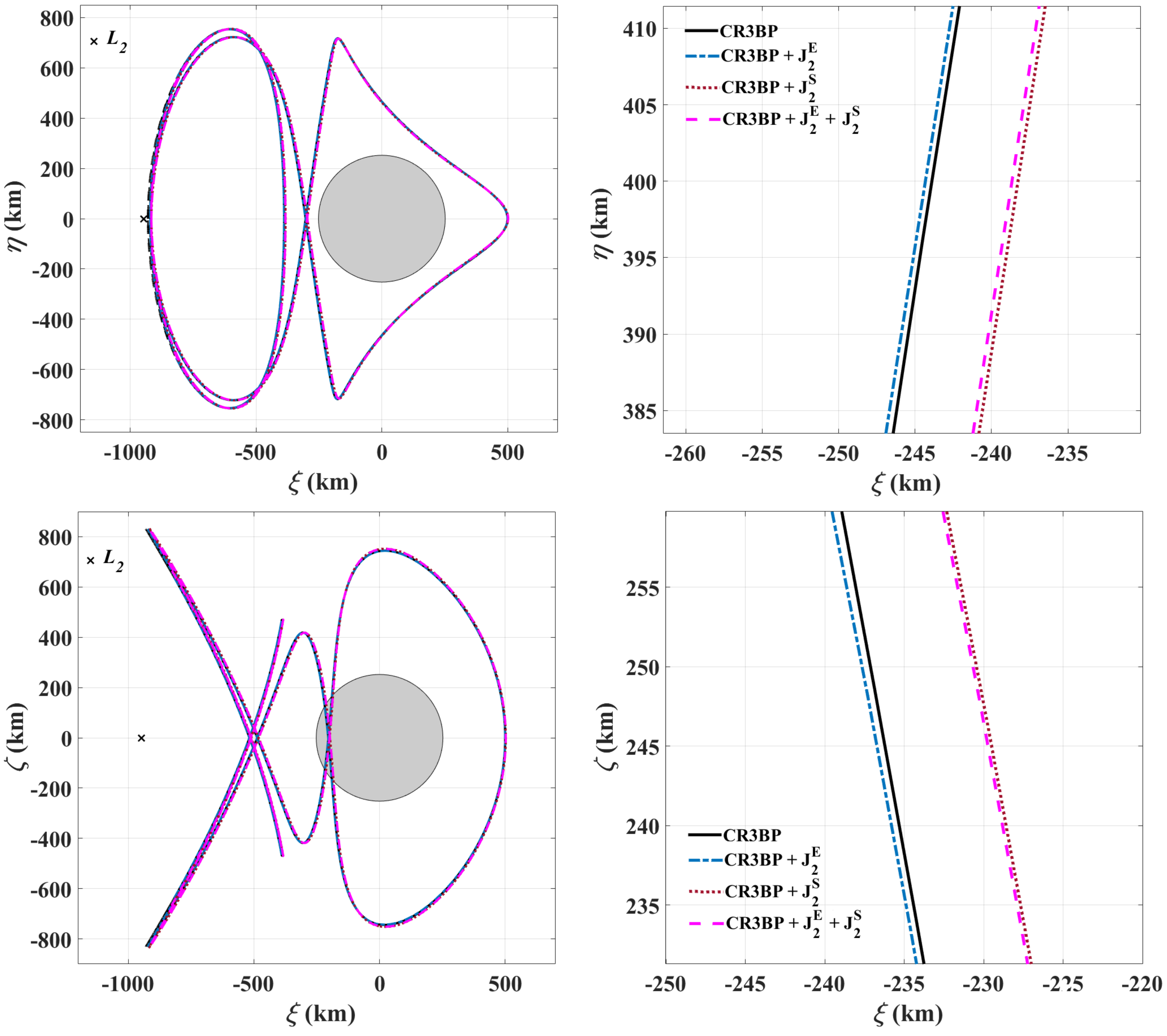}
\caption{Left: planar projections of the heteroclinic connection type D in the three $J_2$-perturbed Saturn-Enceladus CR3BPs (Enceladus-centered synodical frame). Right: zoomed views of the left plots.}
\label{fig:TypeD_Pert}
\end{figure}

At the adopted energy discretisation (100 halo orbits in each family), four heteroclinic connections have been identified in the unperturbed CR3BP (see Fig.~\ref{fig:Heteros_CR3BP}) and have been reproduced in the three perturbed models. They are a Northern $L_1$ halo to a Northern $L_2$ halo transfer (type A), a transfer from a Southern $L_2$ halo to a Northern $L_1$ halo  (type B), a connection from a Northern $L_1$ halo to a Southern $L_1$ halo (type C) and a trajectory from a Southern $L_2$ halo to a Northern $L_2$ halo  (type D).  In all cases, the velocity and position errors at the patch point on $\Sigma$ are smaller than 1 m/s and 1 km, respectively, meaning that these trajectories are approximately continuous both in position and in velocity (the latter implying that they are essentially manoeuver-free). Figures~\ref{fig:TypeA_Pert}-\ref{fig:TypeD_Pert} show projections on two coordinate planes of the perturbed solutions of type A to D according to the three dynamical models and offer a comparison with the unperturbed orbits in zoomed views of the same projections. Tables~\ref{tab:CJ} and \ref{tab:ToF} provide the Jacobi constants and the halo-to-halo transfer times for each connection.     
\begin{table}[h!]
\caption{Jacobi constants of the heteroclinic connections computed in the unperturbed Saturn-Enceladus CR3BP and in the three $J_2$-perturbed models.}
\label{tab:CJ}
\begin{center}
\begin{tabular}{lcccc}
 \hline
Model                              & Type A             & Type B       &  Type C      &   Type D  \\ 
\hline
CR3BP                              & $3.000118$          & $3.000118$  &  $3.000072$  &   $3.000072$ \\
CR3BP + $J^{E}_2$                 & $3.000119$          & $3.000119$  &  $3.000073$  &   $3.000073$ \\
CR3BP + $J^{S}_2$                  & $3.002731$          & $3.002730$  &  $3.002684$  &   $3.002684$ \\
CR3BP + $J^{E}_2$ + $J^{S}_2$     & $3.002731$          & $3.002731$  &  $3.002684$  &   $3.002684$ \\ 
 \hline
\end{tabular}
\end{center}
\end{table}
\begin{table}[h!]
\caption{Halo-to-halo transfer times over the heteroclinic connections computed in the unperturbed Saturn-Enceladus CR3BP and in the three $J_2$-perturbed models.}
\label{tab:ToF}
\begin{center}
\begin{tabular}{lcccc}
\hline
Model                              & Type A   & Type B   &  Type C      &  Type D \\ 
																	 & (hour)   & (hour)   &  (hour)      & (hour)   \\ 
\hline
CR3BP                              & $49.8$          & $39.4$           &  $57.6$  &   $57.6$ \\
CR3BP + $J^{E}_2$                 & $49.7$          & $39.4$           &  $58.0$  &   $58.0$ \\
CR3BP + $J^{S}_2$                  & $49.6$          & $39.4$           &  $58.2$  &   $58.2$ \\
CR3BP + $J^{E}_2$ + $J^{S}_2$     & $49.5$          & $39.3$           &  $58.2$  &   $58.2$ \\ 
 \hline
\end{tabular}
\end{center}
\end{table}

\section{Performance analysis}
\label{sec:obs}
\begin{figure}[h!]
\centering
\includegraphics[scale = 0.2]{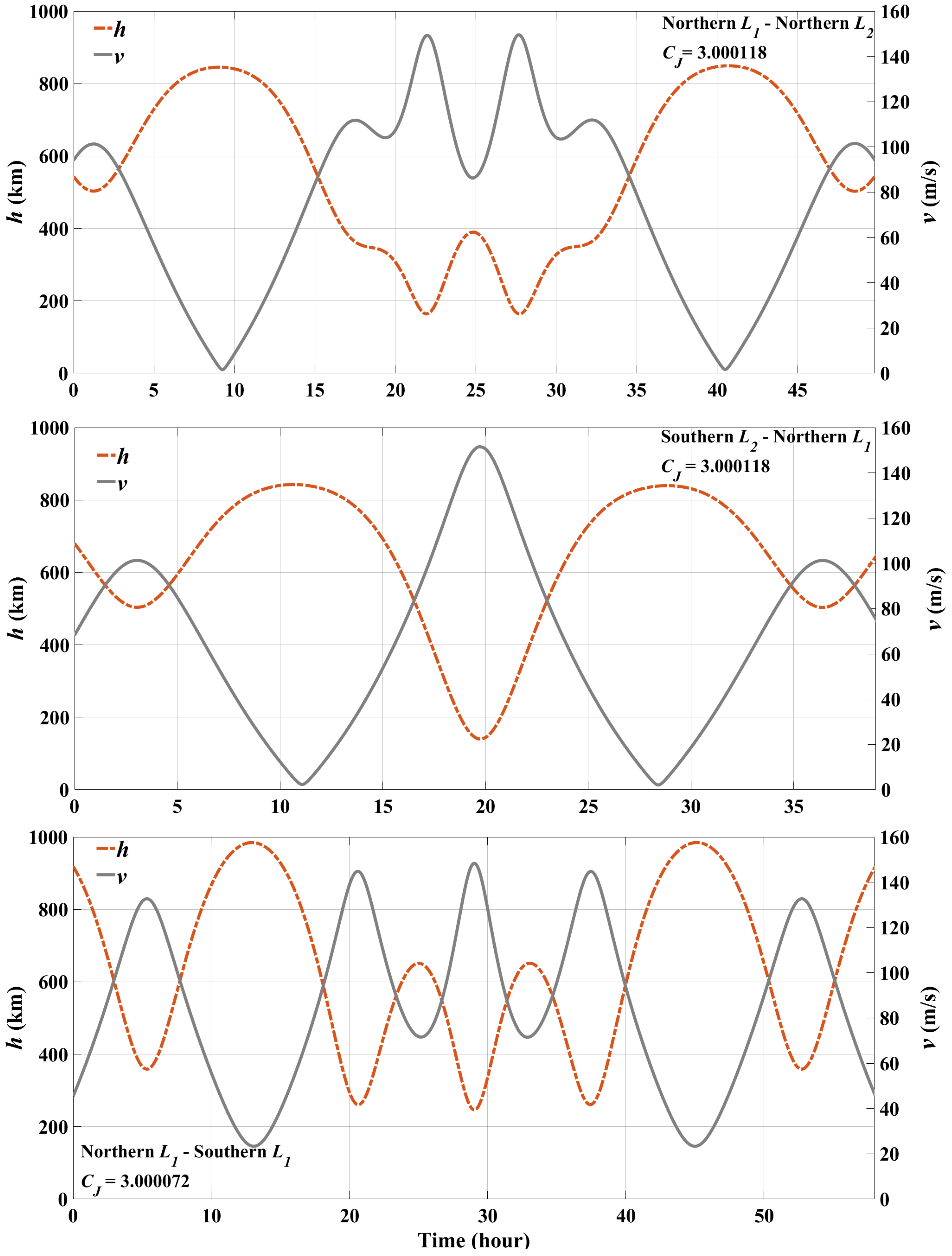}
\caption{Time history of altitude $h$ and velocity $v$ relative to Enceladus over the heteroclinic connections of type A (top), B (middle), and C-D (bottom).}
\label{fig:History_h_v}  
\end{figure}

Figure~\ref{fig:History_h_v} shows the time history of the altitude $h$ above the surface of Enceladus and the magnitude $v$ of the velocity of the spacecraft relative to an Enceladus-centered reference frame with fixed axes for the heteroclinic connections A to D obtained in the unperturbed CR3BP. 
With velocities ranging from $0.08$ m/s at their maximum distance from Enceladus' surface ($\sim$1000 km) to $150$ m/s at their closest approach with the moon ($\sim$200 km), these connecting transfers are extremely convenient in the framework of an \textit{in situ} mission. 

\begin{figure}[h!]
\centering
\includegraphics[scale = 0.40]{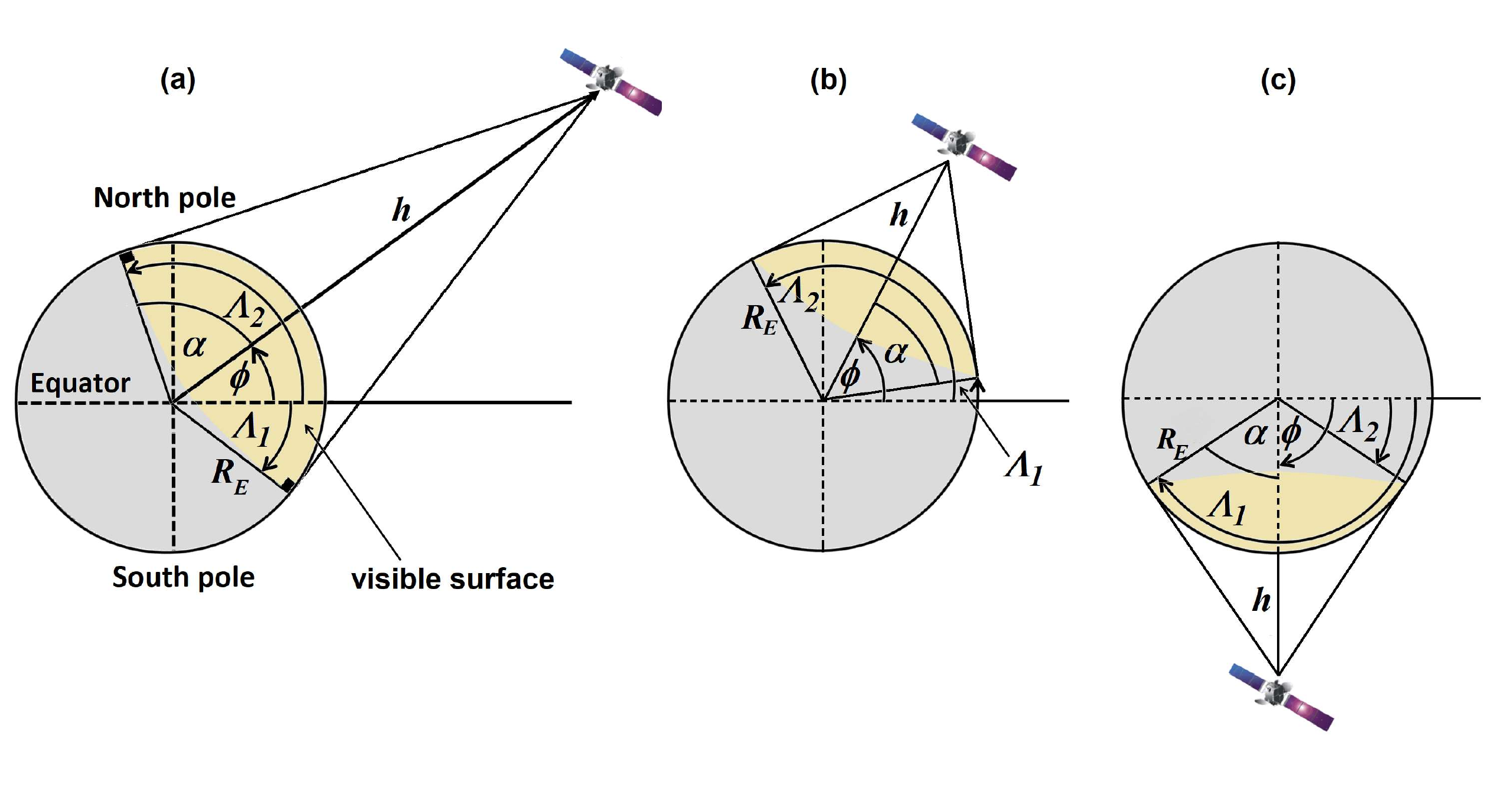}
\caption{The instantaneous coverage parameters $\phi$, $\alpha$, $h$, $\Lambda_1$ and $\Lambda_2$: in (a) the spacecraft has access to the equator and to the north pole, in (b) to the north pole, in (c) to the south pole.}
\label{fig:Cov_Param}  
\end{figure}

\begin{figure}[h!]
\centering
\includegraphics[scale = 0.2]{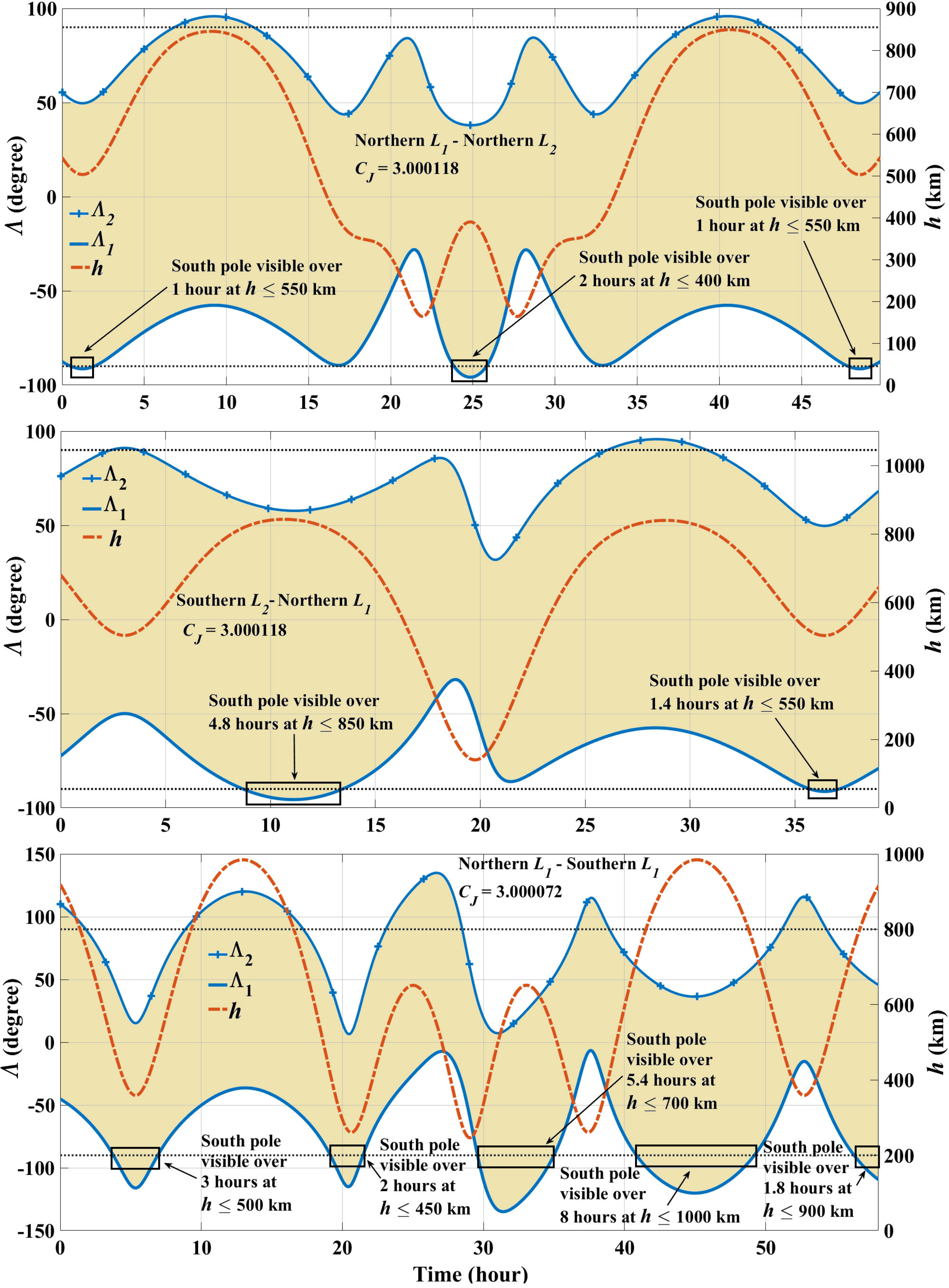}
\caption{Time history of $\Lambda_1$, $\Lambda_2$ and $h$ for the heteroclinic connections of types A (top), B (middle), and C-D (bottom) in the unperturbed CR3BP (from \cite{Fantino:2020}). }
\label{fig:Lambda_old}
\end{figure}

With the objective of quantifying the instantaneous coverage of the surface of the moon, the small difference between the polar and equatorial radii (approximately 4 km) has been neglected and two angles called $\Lambda_1$ and $\Lambda_2$ have been introduced (see Fig.~\ref{fig:Cov_Param}). They represent the limits of the central angle of coverage of amplitude $2\alpha$ and are measured positively northwards from the equator. The angle $\alpha$ depends on the radius $R_E$ of Enceladus and the altitude $h$ of the spacecraft through
\begin{equation}
\alpha = \cos^{-1}\left(\frac{R_E}{R_E+h}\right).
\end{equation}
If $\phi$ denotes the latitude of the spacecraft, then $\Lambda_1$ and $\Lambda_2$ are defined as
\begin{eqnarray}
\Lambda_1 & = & \phi - \alpha, \\
\Lambda_2 & = & \phi + \alpha.
\label{eq:3}
\end{eqnarray}
For instance, the spacecraft has access to the equator, the north pole or the south pole when the interval $\left[\Lambda_1,\Lambda_2\right]$ includes $0^{\circ}$, $90^{\circ}$ or $-90^{\circ}$, respectively. The time history of $\Lambda_1$ and $\Lambda_2$ for the transfers of Fig.~\ref{fig:Heteros_CR3BP} is shown in Fig.~\ref{fig:Lambda_old}, in which the distance between the curves of $\Lambda_1$ and $\Lambda_2$ represents the instantaneous amplitude of coverage, whereas the two dashed lines indicate the latitudes of the poles. Note that the results for heteroclinic connections of type C and D are identical because the time history of $h$ and latitude $\Phi$ are the same, hence they are reported only once. 
These trajectories allow to observe both polar regions of Enceladus. In particular, the south pole is visible during approximately $4$ hours in a type A transfer, during approximately 6 hours in a trajectory of type B and during $21$ hours in the case of transfers of types C and D.

\begin{figure}[h!]
\centering
\includegraphics[scale = 0.19]{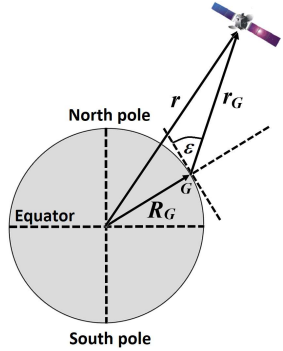}
\caption{Definition of local horizon for a point G on the surface of Enceladus and the corresponding elevation $\varepsilon$ of the spacecraft.}
\label{fig:tau_def} 
\end{figure}

\begin{figure}[h!]
\centering
\includegraphics[scale = 0.18]{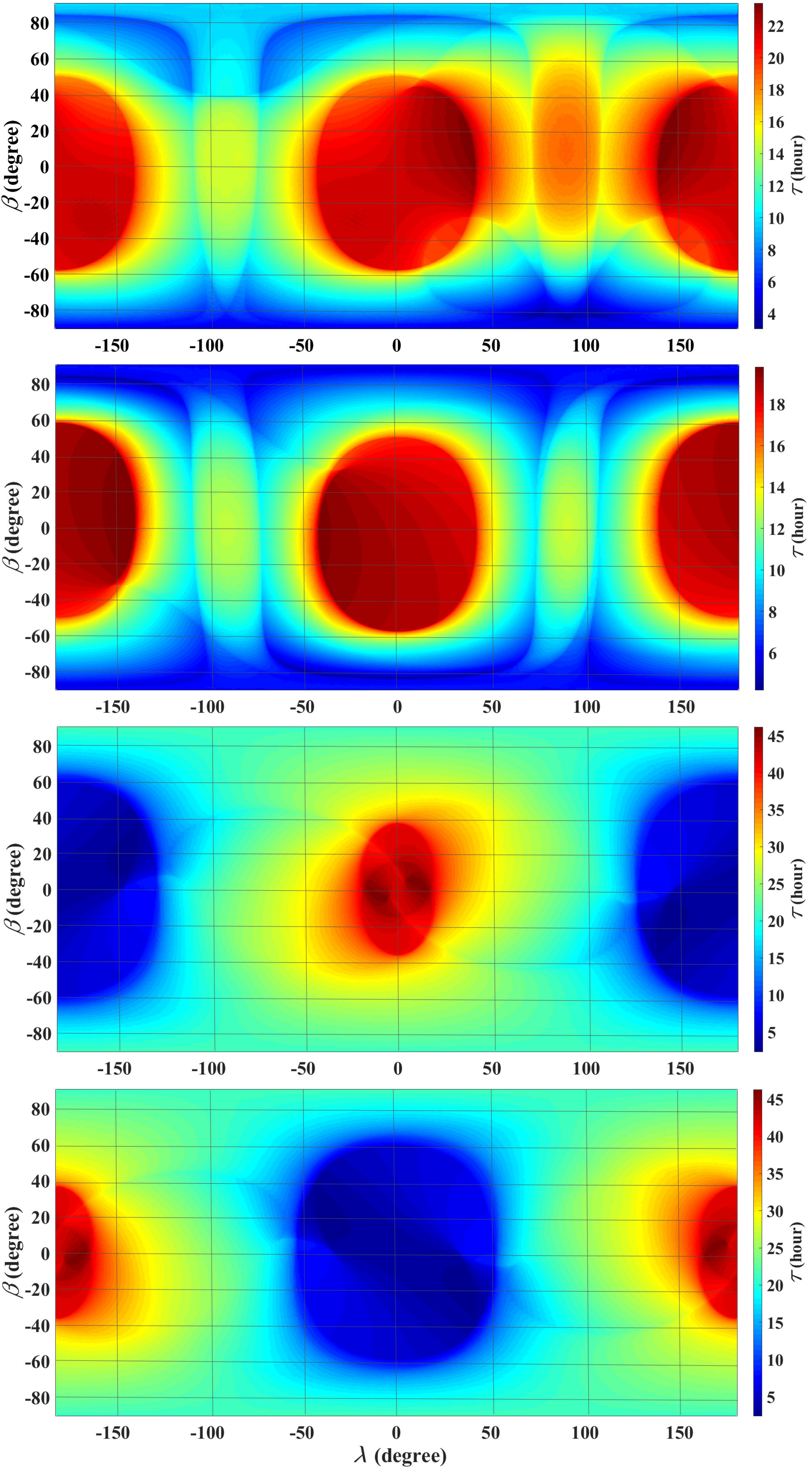}
\caption{Geographical maps of total time of overflight for the heteroclinic connections of the unperturbed CR3BP: types A to D from top to bottom.}
\label{fig:tau_old}  
\end{figure}

We define the total time of overflight $\tau$  as the total access time of a specific surface point over a given transfer. A surface point is visible from the spacecraft when the elevation angle $\varepsilon$ of the latter on the local horizon is positive ($\varepsilon \geq 0$, see Fig.~\ref{fig:tau_def}). Discretising the time of flight on each trajectory in $N$ intervals of duration $\delta t$ and assigning to each an elementary time of overflight $\delta \tau_i$ ($i$ = 1,2,...,$N$),
\begin{equation}
\label{eq:3}
\delta\tau_i =
\begin{cases}
\delta t & \text{if } \epsilon_i \geq 0
\\
0 & \text{otherwise,} 
\end{cases}
\end{equation}
allow to approximate the total time of overflight $\tau$ at the given location as
\begin{equation}
\label{eq:4}
\tau = \sum_{i=1}^{N}\delta\tau_i.
\end{equation}
The computation of $\tau$ has been carried out with $N$ = 1000 and in the Enceladus-centered synodical frame, in this way taking into account the effect of the axial rotation of the primaries (synchronous with their orbital motion).
Figure~\ref{fig:tau_old} illustrates the total time of overflight of Enceladus for the unperturbed heteroclinic connections A to D in the form of geographical maps  of $\tau$. The maps were obtained by discretising the surface of Enceladus at intervals of $0.01$ radians in longitude ($\lambda$) and latitude ($\beta$).     

\begin{figure}[h!]
\centering
\includegraphics[scale = 0.22]{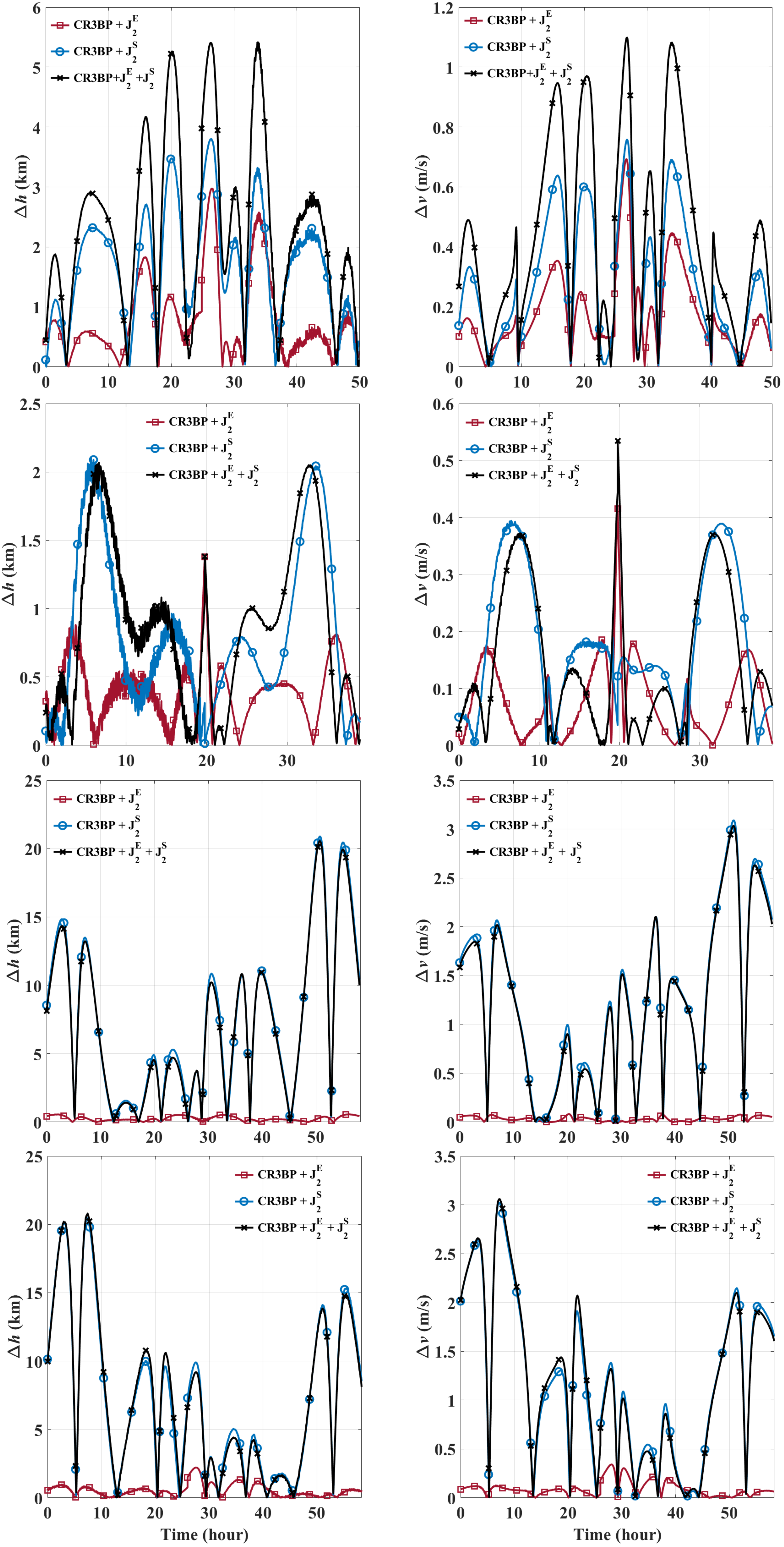}
\caption{Difference in the time histories of altitude and inertial velocity over the four heteroclinic connections (from Type A at the top to type D at the bottom) between the unperturbed CR3BP solutions and each of the three $J_2$-perturbed trajectories.}
\label{fig:h_v_diff}  
\end{figure}

\begin{figure}[h!]
\centering
\includegraphics[scale = 0.22]{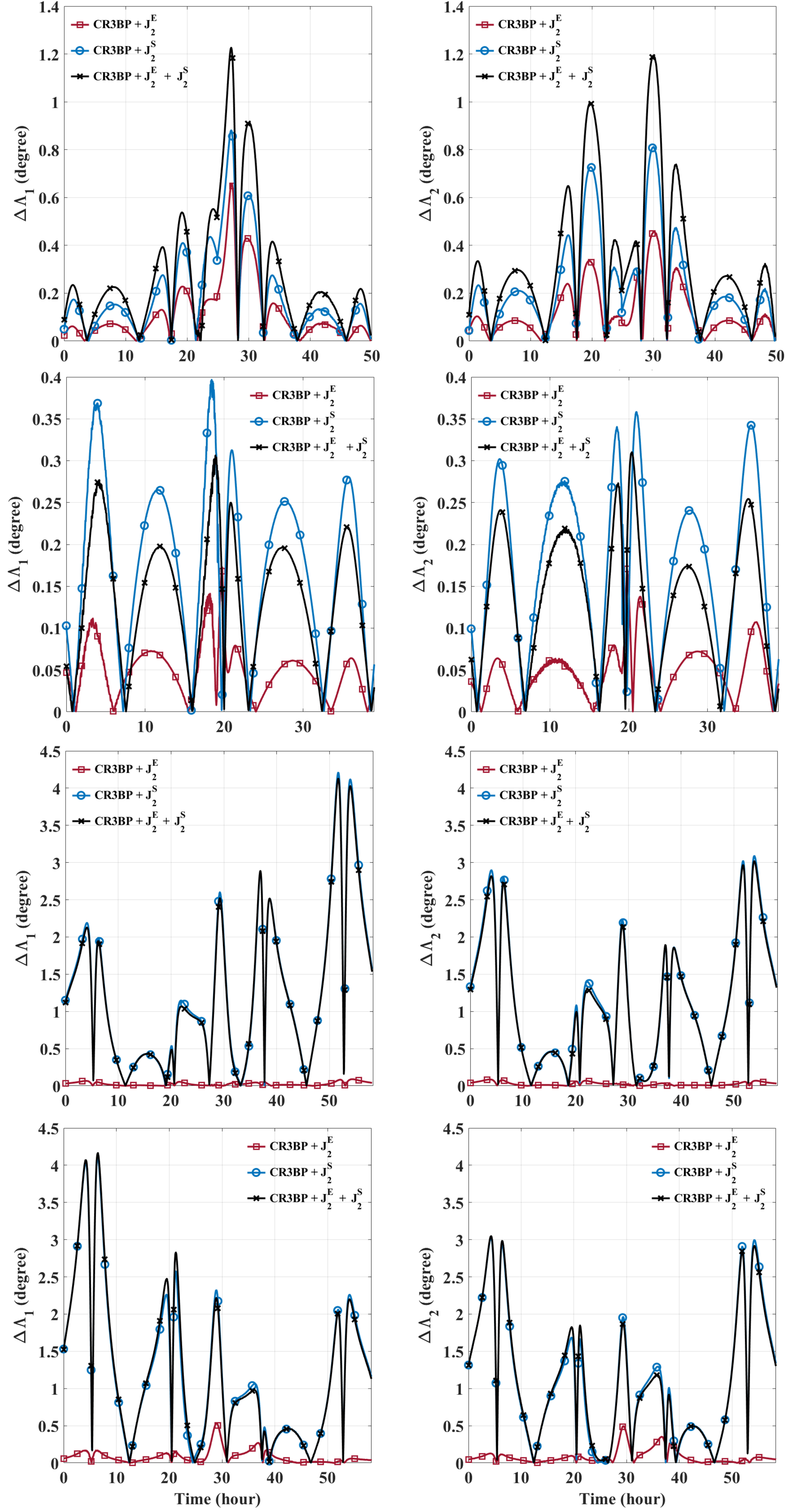}
\caption{Difference in the time histories of the coverage parameters $\Lambda_1$ and $\Lambda_2$ over the four heteroclinic connections (from Type A at the top to type D at the bottom) between the unperturbed CR3BP solution and each of the three $J_2$-perturbed trajectories.}
\label{fig:lambda_diff} 
\end{figure}

\begin{figure}[h!]
\centering
\includegraphics[scale = 0.2]{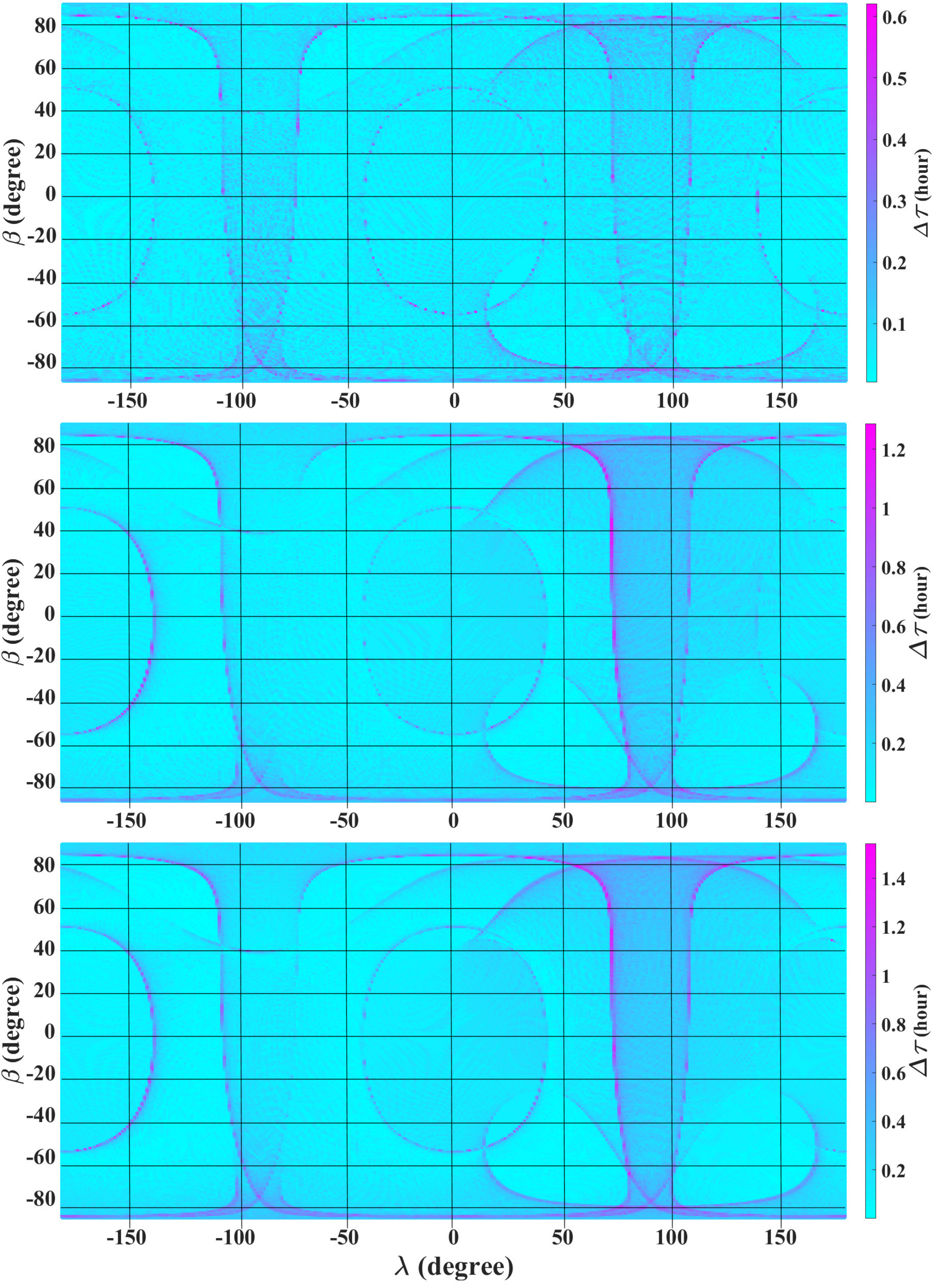}
\caption{Difference in total time of overflight between the unperturbed CR3BP solution of type A and each of the corresponding $J_2$-perturbed trajectories (top: CR3BP + $J^{E}_2$, middle: CR3BP + $J^{S}_2$, bottom: CR3BP +  $J^{E}_2$ + $J^{S}_2$).}
\label{fig:tau_diff_A}   
\end{figure}

\begin{figure}[h!]
\centering
\includegraphics[scale = 0.2]{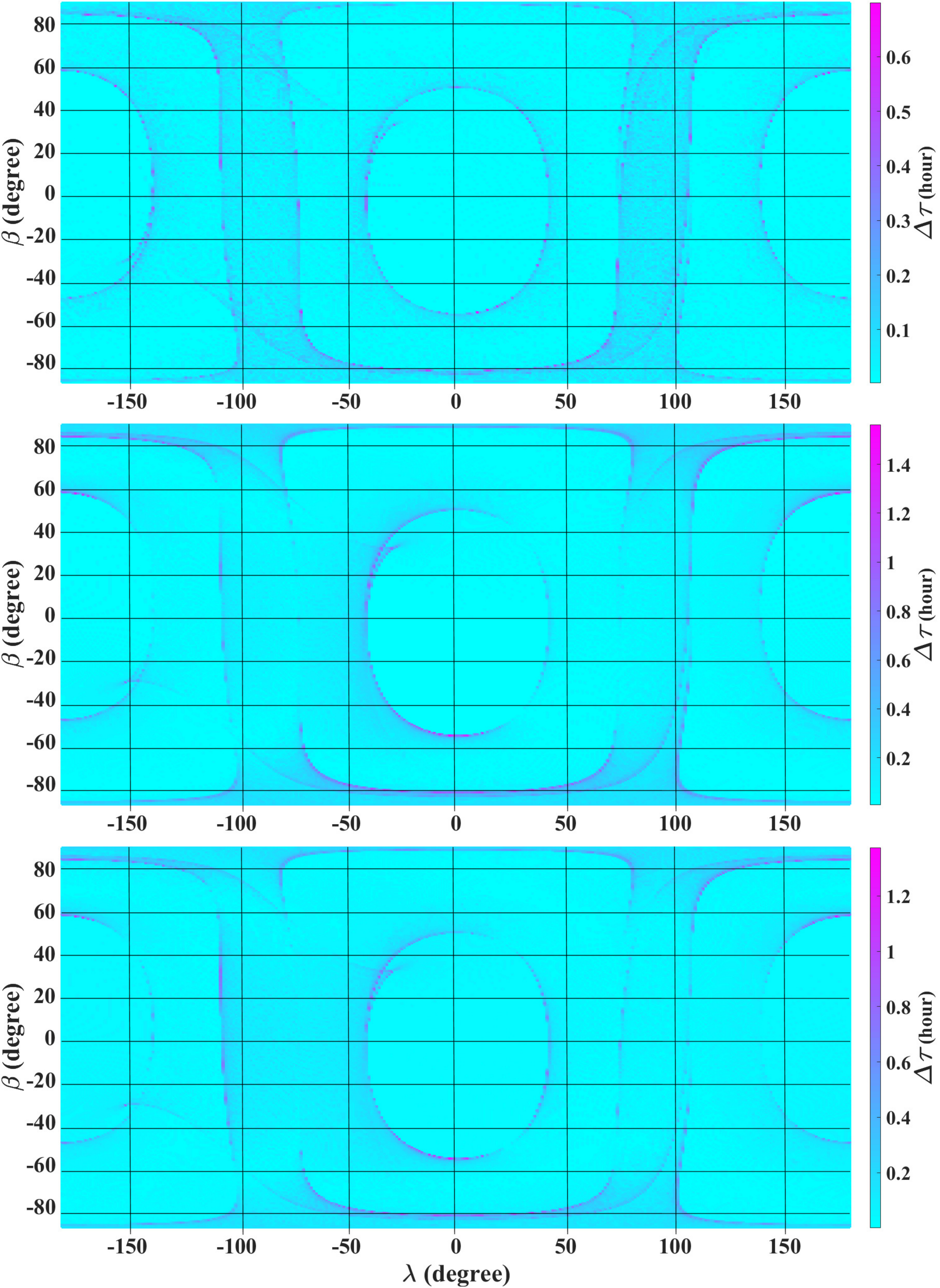}
\caption{Difference in total time of overflight between the unperturbed CR3BP solution of type B and each of the corresponding $J_2$-perturbed trajectories (top: CR3BP + $J^{E}_2$, middle: CR3BP + $J^{S}_2$, bottom: CR3BP +  $J^{E}_2$ + $J^{S}_2$).}
\label{fig:tau_diff_B}   
\end{figure}

\begin{figure}[h!]
\centering
\includegraphics[scale = 0.2]{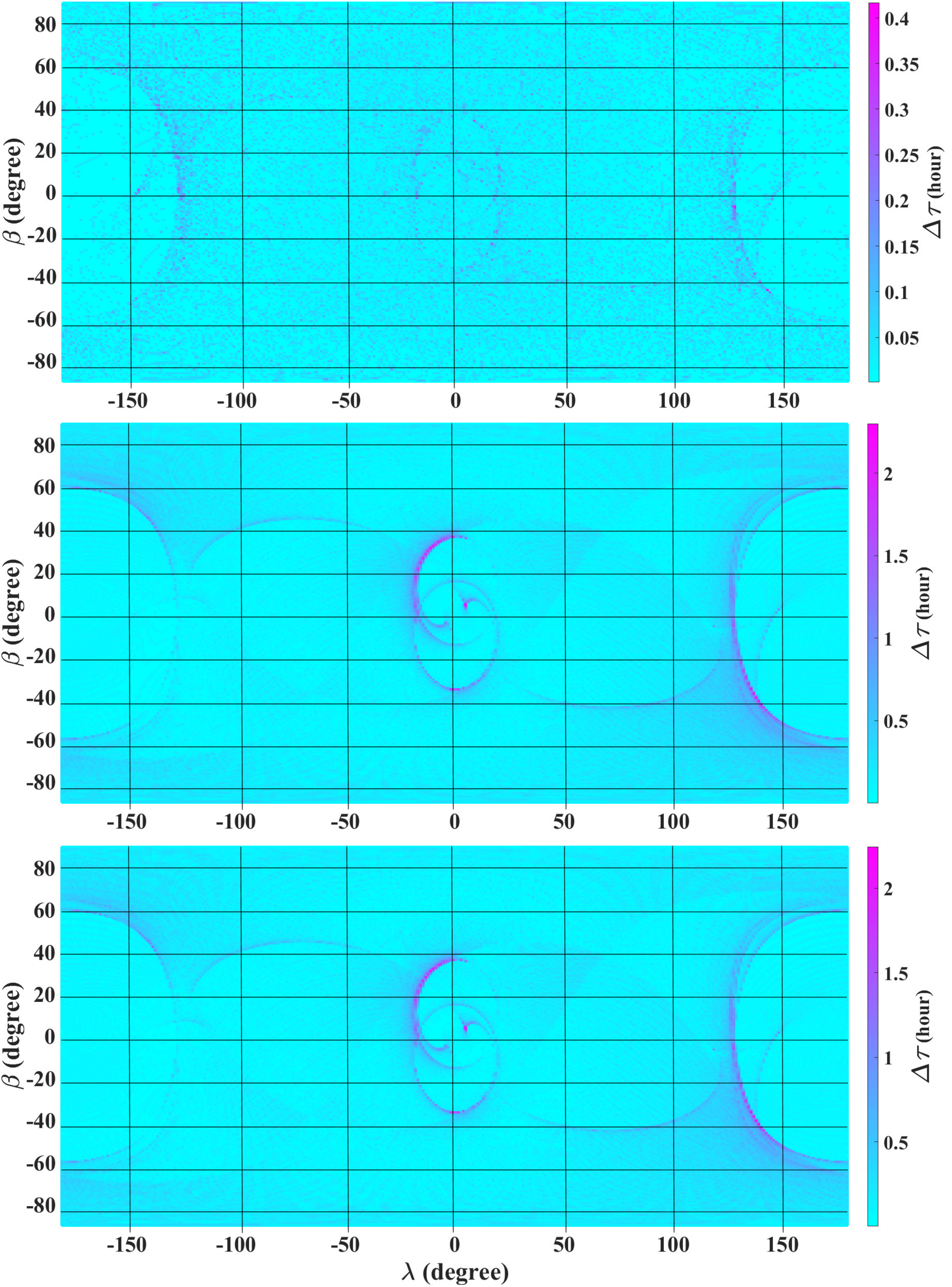}
\caption{Difference in total time of overflight between the unperturbed CR3BP solution of type C and each of the corresponding $J_2$-perturbed trajectories (top: CR3BP + $J^{E}_2$, middle: CR3BP + $J^{S}_2$, bottom: CR3BP +  $J^{E}_2$ + $J^{S}_2$).}
\label{tau_diff_C}   
\end{figure}

\begin{figure}[h!]
\centering
\includegraphics[scale = 0.2]{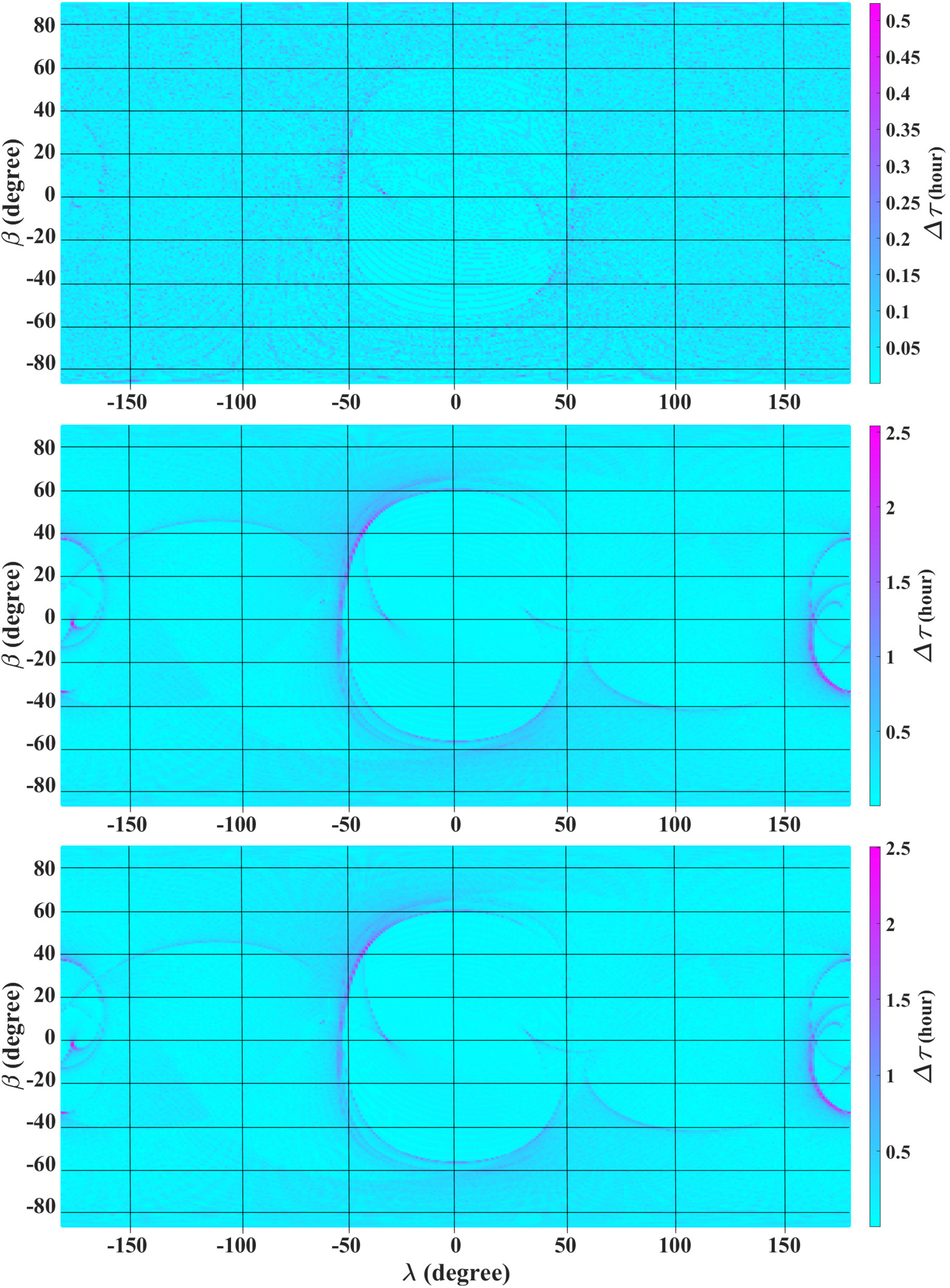}
\caption{Difference in total time of overflight between the unperturbed CR3BP solution of type D and each of the corresponding $J_2$-perturbed trajectories (top: CR3BP + $J^{E}_2$, middle: CR3BP + $J^{S}_2$, bottom: CR3BP +  $J^{E}_2$ + $J^{S}_2$).}
\label{fig:tau_diff_D}  
\end{figure}

The observational performance of the $J_2$-perturbed heteroclinic connections of Figs.~\ref{fig:TypeA_Pert} to \ref{fig:TypeD_Pert} has been compared with that of the corresponding unperturbed solutions. Since the equivalent of Figs.~\ref{fig:History_h_v}-\ref{fig:tau_old} do not allow to appreciate the change with respect to the unperturbed case, we display the point-by-point difference between corresponding results (Fig.~\ref{fig:h_v_diff} for the time history of $h$ and $v$, Fig.~\ref{fig:lambda_diff} for $\Lambda_1$ and $\Lambda_2$, Figs.~\ref{fig:tau_diff_A} to \ref{fig:tau_diff_D} for $\tau$). In assuming that the time coordinate is the same between corresponding evolutions of the same parameter, an approximation is introduced, justified by the similarity in the halo-to-halo transfer times between unperturbed and perturbed solutions (see Table~\ref{tab:ToF}). The error that is introduced, although small, should not be forgotten when interpreting the results.

\section{Discussion}
\label{sec:disc}
The inclusion of the perturbation due to $J_2$ of Saturn and Enceladus in the equations of motion of the spacecraft introduces small differences in the heteroclinic connections between halo orbits computed in the unperturbed CR3BP. The halo orbits themselves have dynamical substitutes in all three versions of the perturbed model.
The orbit-to-orbit transfer times are not significantly altered and this has allowed a point-by-point comparison of the performance parameters. 
The analysis of the individual effects of the oblateness of Saturn and Enceladus shows that the former exerts a stronger perturbation than the latter: a few degrees versus fractions of a degree in the instantaneous coverage angles, 10-20 km versus 1-2 km in the altitude, one hour versus a few minutes in the total access times of the surface of the moon. The higher intensity of the effect of Saturn varies among the four solutions, being more relevant in the case of trajectories of type C and D
which reach larger distances from Enceladus and maintain them for longer times. The $J_2$ effect of Enceladus is more noticeable when the spacecraft approaches the surface of the moon.

\section{Conclusions}
\label{sec:conc}
The present contribution extends previous investigations of low-energy trajectories around Enceladus to include the effects of the oblateness of the two primaries, i.e., Saturn and Enceladus. By applying and unifying the theoretical developments of other authors (most noticeably, \cite{Bury:2018, Arredondo:2012, Abouelmagd:2012, Abouelmagd:2015}), we have described the features of the Saturn-Enceladus circular restricted three-body problem (CR3BP) with oblate primaries, including the positions of the equilibria and the dynamical susbstitutes of the halo orbits (over a wide energy range) and their stable and unstable hyperbolic invariant manifolds. Also the heteroclinic connections identified by \cite{Fantino:2020} in the unperturbed CR3BP by intersecting stable and unstable HIMs of two halo orbits of $L_1$ and $L_2$ at a certain surface of section have been refined in the perturbed model by introducing the oblateness of the two primaries separately and in combination, thus gaining a quantitative insight into the individual effects.  
The reason for the interest in these trajectories resides in their high inclination, close approaches to the moon and negligible cost (discontinuities in position and velocity at the patch point are at the level of 1 km and 1 m/s, respectively) and in the periodic character of departure and arrival orbits which can be used as spaceports between consecutive observation flights.

The study shows that the shape and energy of the solutions change, although not significantly, when the oblateness of the primaries is taken into account. The analysis of the observational performance also shows small differences in the geometrical and kinematical coverage parameters, with $J_2$ of Saturn playing a larger role than the oblateness of Enceladus, especially for the trajectories that reach the edges of the Hill sphere of the moon (i.e., the distance of $L_1$ and $L_2$ from the center of the body, approximately 950 km).

In general, the results substantially confirm the features of the solutions obtained in the unperturbed CR3BP. When the complete model (including the oblateness of both primaries) is considered, the entire surface of Enceladus is still visible from the spacecraft, uninterrupted windows of access to the southern polar region exist and extend over several hours, the specific duration depending on altitude.
For example, in the heteroclinic trajectory that connects the Northern halo of $L_1$ with the Northern halo of $L_2$  at $C_J$ = 3.002731, the south pole is accessible during 2 hours from below 400 km altitude, whereas in the solution connecting a Northern halo with a Southern halo at $L_1$ with $C_J$ = 3.002684 the south pole is visible for over 20 hours distributed along four windows at different altitudes.
The detailed assessment of the time of overflight (defined as the time spent by the spacecraft above the local horizon) of a regular grid of points over the surface has been expressed in the form of geographical color maps. These maps show that the local cumulative visibility is never shorter than 4 hours (polar regions) with peaks of 20 or even 40 hours for wide equatorial bands (up to $\pm$ 60 degrees latitude). 

In conclusion, the connections obtained in the classical (unperturbed) Saturn-Enceladus CR3BP persist upon transition to a more accurate model and their performance features as science orbit remain valid and very appealing, thus suggesting that the simpler model can be used in a preliminary feasibility analysis.
The methodology and the application here exposed can be implemented in the case of other targets of interest exhibiting peculiar features over extended portions of their surface, such as Uranus' moons Titania and Oberon or Jupiter's Europa. 

\section*{Appendix A: Mean motion of a pair of oblate spheroids in a circular orbit perpendicular to their axes of maximum inertia} 
\label{sec:app}
The gravitational potential $V(P)$ of an oblate spheroid of mass $m$ at an external point $P$ is given by 
\begin{equation}
V(P) = \displaystyle \frac{Gm}{r}\left[1 - J_2 \left(\frac{a}{R}\right)^2 P_2(\sin \delta)\right],
\label{eq:V_spheroid}
\end{equation}
where $\delta$ denotes the latitude of $P$ with respect to the equator of the spheroid, $r$ is the distance of $P$ from the center of the body, $R$ is the reference  radius of the spheroid (i.e., the equatorial radius), $J_2$ represents the difference between the principal moments of inertia of the body (a positive quantities for oblate bodies) and $P_2$ is the Legendre polynomial of degree 2,
\begin{equation}
P_2(x) = \displaystyle \frac{3x^2-1}{2}.
\label{eq:Legendre}
\end{equation}
Taking the gradient of the potential and setting $\delta = 0$ yields the acceleration ${\bf a}_P$ of P 
when this is on the equatorial plane of the spheroid:
\begin{equation}
{\bf a}_P = \displaystyle -\frac{Gm}{r^2}\left(1 + \frac{3J_2R^2}{2r^2}\right) {\bf u}_r = -\frac{Gm}{r^2}\left(1 + \frac{\kappa}{r^2}\right) {\bf u}_r,
\label{eq:accP}
\end{equation}
${\bf u}_r$ being the unit vector in the outward radial direction at $P$ and $\kappa = 3J_2R^2/2$. This acceleration accounts for the gravity field of $m$ as a point mass and the effect of the quadrupole moment of its mass distribution.

\begin{figure}[h!]
\centering
\includegraphics[scale = 0.4]{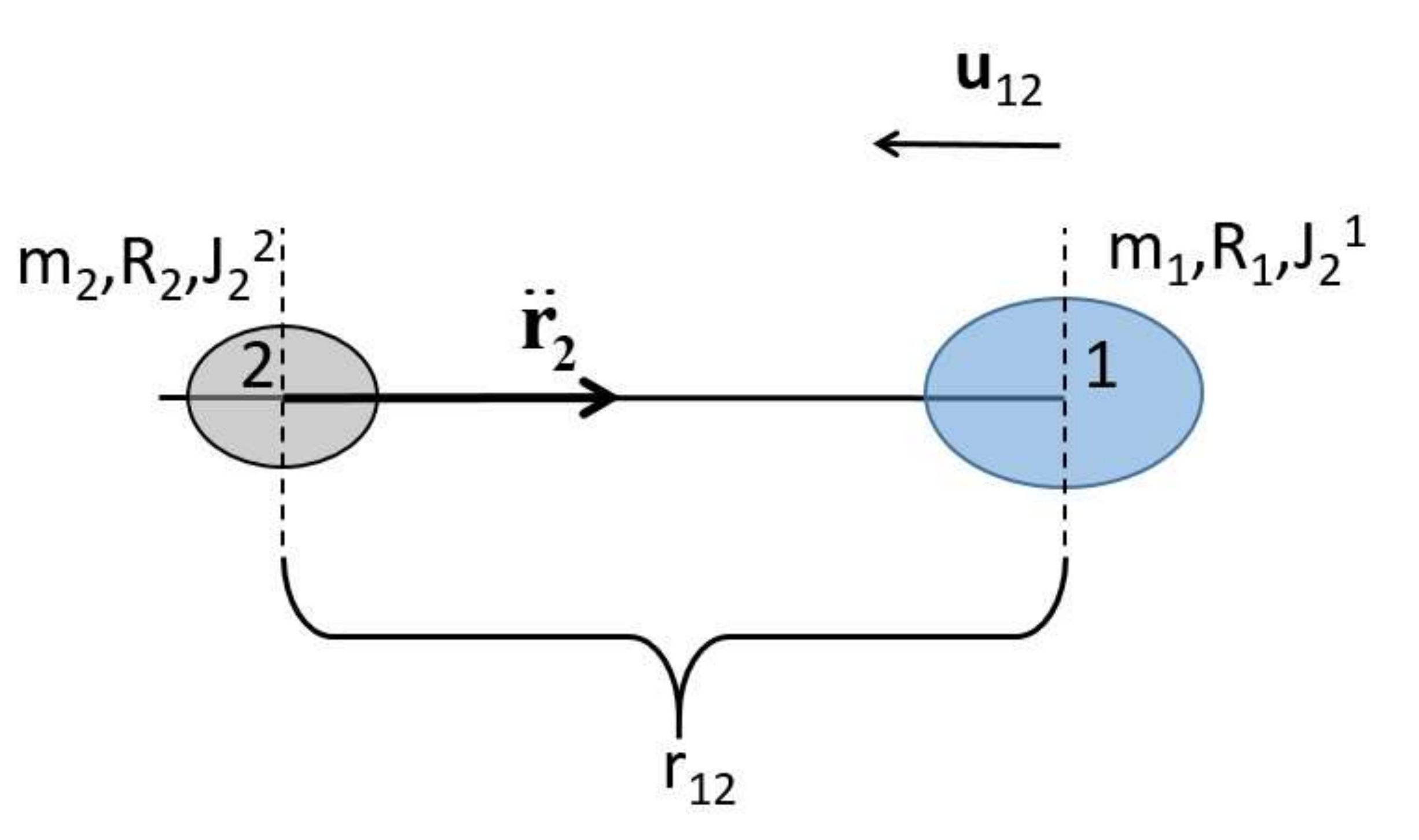}
\caption{Sketch of a system of two oblate homogeneous spheroids with maximum inertia axes parallel to each other.}
\label{fig:app} 
\end{figure}

Let us consider two oblate spheroids in circular orbits perpendicular to their axes of maximum inertia (Fig.~\ref{fig:app}). If ${\bf u}_{12}$ denotes the unit vector from the center of body 1 to the center of body 2, {\bf the acceleration $\ddot{\bf r}_2$ of body 2 is }
\begin{equation}
\ddot{\bf r}_2 = \displaystyle \left[ -\frac{Gm_1}{r_{12}^2}\left(1 + \frac{\kappa_1}{r_{12}^2} + \frac{\kappa_2}{r_{12}^2}\right) + {\rm \it O}\left(\frac{\kappa_i}{r_{12}^2}\right)^2 \right] {\bf u}_{12},
\label{eq:f2}
\end{equation}
$\kappa_i$ being $3J_2^i R_i^2/2$, $i$ = 1,2.
The right-hand side of Eq.~\ref{eq:f2} accounts for:
{\bf \begin{itemize} 
\item the acceleration of point mass 2 due to point mass 1 $\displaystyle \left(-\frac{Gm_1}{r_{12}^2}{\bf u}_{12}\right)$;
\item the effect of the quadrupole moment of the mass distribution of body 1 over point mass 2 $\left(\displaystyle -\frac{Gm_1\kappa_1}{r_{12}^4}{\bf u}_{12}\right)$ (see also the second term of Eq.~\ref{eq:accP});
\item the acceleration caused by point mass 1 over the quadrupole moment of the mass distribution of body 2 $\displaystyle \left(-\frac{Gm_1\kappa_2}{r_{12}^4}{\bf u}_{12}\right)$. This term has been obtained by application of the third law of dynamics: as a matter of fact, the corresponding force $\displaystyle \left(-m_2\frac{Gm_1\kappa_2}{r_{12}^4}{\bf u}_{12}\right)$is equal in magnitude and opposite in direction to the attraction that the quadrupole moment of the mass distribution of body 2 exerts on point mass 1 $\displaystyle \left(m_1\frac{Gm_2\kappa_2}{r_{12}^4}{\bf u}_{12}\right)$ (see also the second term of Eq.~\ref{eq:accP}).
\end{itemize} }
The effect of the quadrupole moment of the mass distribution of body 1 over the quadrupole moment of the mass distribution of body 2 is an infinitesimal of order higher than  
$(\kappa_i/r_{12}^2)^2$ and, for the application considered in this work, it can be neglected because it less  important than the 
error due to assuming circular equatorial orbits for the primaries.
Therefore, the acceleration $\ddot{\bf r}_2$ of body 2 can be written as
\begin{equation}
\ddot{\bf r}_2 = \displaystyle -\frac{Gm_1}{r_{12}^2}\left(1 + \frac{\kappa_1+\kappa_1}{r_{12}^2}\right) {\bf u}_{12},
\label{eq:acc2}
\end{equation}
and, similarly, the acceleration $\ddot{\bf r}_1$ of body 1 is
\begin{equation}
\ddot{\bf r}_1 = \displaystyle +\frac{Gm_2}{r_{12}^2}\left(1 + \frac{\kappa_1+\kappa_2}{r_{12}^2}\right) {\bf u}_{12},
\label{eq:acc1}
\end{equation}
Eventually, Kepler's third law applied to the relative motion 
\begin{equation}
\ddot{\bf r}_2-\ddot{\bf r}_1 = \displaystyle -\frac{G(m_1+m_2)}{r_{12}^2}\left(1 + \frac{\kappa_1+\kappa_1}{r_{12}^2}\right) {\bf u}_{12}
\label{eq:accrel}
\end{equation}
yields the mean motion $n$ of the system:
\begin{equation}
n= \displaystyle \sqrt{\frac{G(m_1+m_2)}{r_{12}^3}\left(1 + \frac{\kappa_1+\kappa_1}{r_{12}^2}\right)}.
\label{eq:n_app}
\end{equation}

\section{Acknowledgments}
This investigation has been supported by Khalifa University of Science and Technology's internal grants FSU-2018-07 and CIRA-2018-85. The authors thank the editor and the reviewer for their constructive and valuable comments and R. Flores for his timely support and enlightening thoughts.


\bibliography{bib_J2_new_R1}

\end{document}